\documentclass[conference,a4paper]{IEEEtran}
\IEEEoverridecommandlockouts
\usepackage{amsmath,amssymb,amsthm}
\usepackage{graphicx}
\usepackage{booktabs}
\usepackage{array}
\newcolumntype{L}[1]{>{\raggedright\arraybackslash}p{#1}}
\usepackage{cite}
\usepackage{tikz}
\usetikzlibrary{arrows.meta,calc,decorations.pathreplacing,positioning}
\definecolor{cUtil}{HTML}{2A6FB8}
\definecolor{cIsl}{HTML}{E0861B}
\definecolor{cHold}{HTML}{139B75}
\definecolor{cRef}{HTML}{A8327F}
\definecolor{cAtk}{HTML}{7A1414}
\definecolor{cInk}{HTML}{1A1A1A}
\definecolor{cMuted}{HTML}{707070}
\tikzset{
  >={Stealth[length=3.6pt,width=2.8pt]},
  paper/.style={font=\footnotesize, line cap=round, line join=round},
  sig/.style={->, draw=cInk!85, line width=0.5pt},
  atk/.style={->, draw=cAtk, line width=0.9pt},
}
\usepackage[hidelinks]{hyperref}

\newtheorem{theorem}{Theorem}
\newtheorem{lemma}{Lemma}
\newtheorem{proposition}{Proposition}
\newtheorem{corollary}{Corollary}
\theoremstyle{definition}
\newtheorem{definition}{Definition}
\newtheorem{assumption}{Assumption}

\newcommand{\Ev}{\mathcal E}
\newcommand{\Es}{\mathcal E^\star}
\newcommand{\KL}{D_{\mathrm{KL}}}
\newcommand{\tc}{\tau^{\mathrm c}}
\newcommand{\gb}{\bar\gamma}
\newcommand{\N}{\mathcal N}
\newcommand{\pth}{\tilde\theta}
\newcommand{\pnu}{\tilde\nu}
\newcommand{\tref}{\theta^{\mathrm{ref}}}
\newcommand{\R}{\mathbb R}
\newcommand{\Nat}{\mathbb N}
\newcommand{\Z}{\mathbb Z}
\newcommand{\Sym}{\mathbb S}
\newcommand{\Kw}{\mathcal K}
\newcommand{\A}{\mathcal A}
\newcommand{\Dset}{\mathcal D}
\newcommand{\Jset}{\mathcal J}
\newcommand{\Yp}{\mathcal Y_-}
\newcommand{\Pb}{\mathbb P}
\newcommand{\Ex}{\mathbb E}
\newcommand{\Ric}{\mathrm{Ric}}
\newcommand{\inwords}{\noindent\emph{Interpretation.}\ }

\begin{document}

\title{An Inspection Game for Monitor Design Against\\ GNSS Timing Attacks in Synchronous Islanding}

\author{\IEEEauthorblockN{Gregory Provan}
\IEEEauthorblockA{School of Computer Science and IT\\
University College Cork, Ireland}
\and
\IEEEauthorblockN{Quanyan Zhu}
\IEEEauthorblockA{Department of Electrical and Computer Engineering\\
New York University, USA}}

\maketitle

\begin{abstract}
Phasor measurement units (PMUs) take their time from GNSS, so a spoofer that slowly drags a receiver clock rotates the reported phase without touching the power waveform. The threat is acute in synchronous islanding, where a controller keeps an islanded generator in phase with the grid, using one PMU on each side of an open breaker, so that the two can be reconnected at any moment: a differential clock bias of about $0.56$\,ms can make the controller hold the generator $10^\circ$ out of phase while it reports perfect synchronism. We cast the defense as a zero-sum inspection game. The defender, as inspector, chooses its timing monitors and audits the clocks periodically; the attacker, as inspectee, drags the two clocks to an unsafe bias by the next audit; the payoff is the evidence, the Kullback--Leibler divergence between the probability laws of the defender's record under attack and under normal operation. Our central result is that the evidence equals the cost of the cheapest innocent explanation of the attacked record. This single identity predicts which monitors work, turns the attacker's best response into a strictly convex quadratic program, and turns the defender's design into a convex program with a saddle point. Pairing explanations, which bound the evidence from above, with test statistics, which bound it from below, certifies audit periods in which every successful attack leaves at least a threshold of evidence. On a 50\,Hz island, these evidence-based periods, at the matched-test threshold, are up to 8.4 times as long as those of a rule that checks each monitor separately, and the case study favors placing the better oscillator on the utility side of the breaker.
\end{abstract}


\section{Introduction}

A phasor measurement unit (PMU) measures the phase angle of a bus voltage against a reference cosine synthesized from its own clock, and it disciplines that clock to the Global Navigation Satellite System (GNSS). A clock error of $\tau$ seconds therefore rotates the reported angle by $2\pi f_0\tau$ radians, where $f_0$ is the nominal frequency~\cite{c37118}; at $f_0=50$\,Hz, one millisecond is $18^\circ$. A spoofer that captures the receiver can drag its clock and thereby shift the reported phase without touching the power system. Such attacks are inexpensive to mount~\cite{humphreys2008}, have been demonstrated against PMUs~\cite{shepard2012}, and corrupt the applications built on synchrophasors~\cite{zhang2013}.

The consequences are most severe where the reported phase is the controlled variable. In \emph{synchronous islanding}~\cite{best2007,best2011}, a generator that has separated from the grid is held in phase with it, using one PMU on each side of the open breaker, so that the breaker can be closed at any moment. If the two PMU clocks disagree by $\tau_r-\tau_g$ seconds, the controller drives the \emph{reported} phase difference to zero and thereby holds a \emph{true} phase difference of $2\pi f_0(\tau_r-\tau_g)$. Closing the breaker outside a synchronization window can cause a damaging transient. We adopt $10^\circ$, the strictest phase limit of~\cite[Table~5]{ieee1547} (for resources above 1500\,kVA), as the operating threshold of this study; at 50\,Hz it corresponds to a clock disagreement of only $0.56$\,ms.

Without a model of normal operation, neither PMU can distinguish clock manipulation from a suitable physical disturbance. With the breaker open, a drifting clock on the utility side looks like a change of grid frequency, and one on the island side like a change of load; only the statistics of these disturbances make the two accounts distinguishable, and often barely so. Operators therefore audit the clocks periodically against an independent time source, and install \emph{monitors} that keep their own time: holdover oscillators and, where available, a trusted phase reference. This raises a design question with a precise answer: \emph{how far apart may the audits be, so that every attack that succeeds between two audits leaves enough evidence to be detected, and which monitors lengthen that period the most?}

\emph{Approach.} We answer this question with a two-player \emph{inspection game}~\cite{avenhaus2002}, stated in Section~\ref{sec:game} before any physics. The defender, the inspector, moves first: it chooses its monitors and announces them. The attacker, the inspectee, who knows the design and the audit schedule, then chooses how to drag the two clocks between two audits, subject to a rate limit, so that reconnection is unsafe at the next audit. The payoff is the \emph{evidence}, the Kullback--Leibler (KL) divergence between the laws of the defender's record with and without the attack; the attacker wants it small and the defender large. Under the Gaussian model, the evidence determines the detection probability of the likelihood-ratio test matched to a specified attack, which by the Neyman--Pearson lemma is the most powerful test against that attack. Audits are \emph{safe} if every successful attack reveals at least a threshold of evidence. At the matched-test threshold for a chosen false-alarm probability and a $50\%$ detection target, safety is necessary for any detector to meet that target against every successful attack, and sufficient for the test matched to each attack; a single detector for unknown attack profiles requires a composite test, which we leave to future work.

\emph{Key insight.} The defender never observes the attack itself; it observes a record that varies randomly even without an attack. An attack is therefore hard to see exactly when the attacked record could plausibly have come from normal operation. We show that the evidence equals the cost of the \emph{cheapest innocent explanation}: the least unusual account of normal operation (a slightly low grid frequency, a slightly heavy load, a slightly slow oscillator) that reproduces what the defender sees. Each monitor is thus characterized by what it can blame a clock drift on, and the whole analysis follows from this one identity.

\emph{Contributions.}
\begin{enumerate}
\item \emph{Model and game.} A formal inspection game on an exactly discretized model of the island and its controller, with a first-order model of how the clocks enter the PMU reports and of the monitors, in which every quantity is typed by its domain.
\item \emph{Evidence as explanation} (Theorem~\ref{thm:evidence}). The evidence is a quadratic form that does not depend on the controller, and it equals the cost of the cheapest innocent explanation. Explanations bound it from above and test statistics from below, which sorts the monitors into three regimes.
\item \emph{The attacker.} The stealthiest attack is the unique solution of a strictly convex quadratic program. When the rate limit does not bind, it splits the bias between the two clocks in a harmonic ratio, and we characterize when a slow ramp or a late sprint comes close to its value.
\item \emph{The defender.} The design problem is a convex program and the game has a saddle point, so the defender loses nothing by announcing its design; under a stated approximation, and in the case study, a better oscillator belongs on the utility side.
\item \emph{Certified audit periods.} Lower bounds on the safe audit period that hold against every attack. For the case-study designs with a monitor they lie within $0.6\%$ of the optimum, as bracketed by explicit feasible attacks, and at the matched-test threshold they are 1.1 to 8.4 times as long as those of a rule that checks each monitor separately.
\end{enumerate}

\emph{Related work.} Stealthy attacks on control systems have been studied from the viewpoints of detectability and identifiability~\cite{pasqualetti2013} and of the resources an adversary needs~\cite{teixeira2015}. When detectability is measured by the KL divergence, as here, \cite{bai2017,mo2016,kung2017} maximize the damage subject to a bound on the divergence. We solve the converse problem, which is to minimize the divergence subject to causing damage, under the structural constraints of clock manipulation (two clocks, a rate limit and an audit deadline), and we then optimize the monitors against the solution. GNSS spoofing of PMUs has been demonstrated~\cite{shepard2012} and its impact on grid applications analyzed~\cite{zhang2013}; common defenses compare the receiver clock with a local oscillator~\cite{shepard2011,psiaki2016}. We show that fusing these comparisons with the phasor data, and placing the better oscillator correctly, lengthens the evidence-based safe audit period severalfold in our case study. Inspection games~\cite{avenhaus2002} model an inspector who verifies with statistical tests that an inspectee does not violate an agreement, typically requiring that a violation be detected before a critical time, and the Neyman--Pearson lemma is their standard tool. Ours is an inspection game in which the inspector chooses the precision of its instruments rather than the times of its inspections, and the violation is a clock profile constrained by a rate limit. Unlike game-theoretic designs of resilient control~\cite{zhu2015}, it fixes the controller and the alarm policy, which yields a convex formulation with closed-form bounds and special-case solutions; the monitor design is an instance of optimal experimental design~\cite{pukelsheim2006}.

\subsection{Notation and conventions}\label{sec:notation}

\emph{Sets and matrices.} $\Nat=\{1,2,\dots\}$, $\Z_{\ge0}=\{0,1,2,\dots\}$, and $\Kw_N=\{0,1,\dots,N-1\}$ for $N\in\Nat$. Vectors are columns. $\Sym^n$, $\Sym^n_+$ and $\Sym^n_{++}$ are the sets of symmetric, positive semidefinite and positive definite $n\times n$ real matrices, and $A\succeq B$ means $A-B\in\Sym^n_+$. For $A\in\Sym^n_+$ and $x\in\R^n$, $\|x\|_A^2=x^\top Ax$. $I_n$ is the identity, $\mathrm{diag}(\cdot)$ a (block) diagonal matrix, $(x)_+=\max\{x,0\}$, $\lfloor x\rfloor$ the largest integer not above $x$, and $\mathbf 1_S\in\{0,1\}$ equals $1$ if the statement $S$ holds. We set $\inf\emptyset=+\infty$; for suprema of sets of nonnegative periods or indices, which lie in $[0,\infty]$, we set $\sup\emptyset=0$.

\emph{Probability.} All random variables are defined on one probability space, and $\Ex$ is expectation. $\N(\mu,C)$ is the Gaussian law with mean $\mu\in\R^n$ and covariance $C\in\Sym^n_+$; $\Phi:\R\to(0,1)$ is the standard normal distribution function, and $z_p=\Phi^{-1}(p)$ for $p\in(0,1)$. For two laws $\Pb_1,\Pb_0$ on $\R^n$, the KL divergence is $\KL(\Pb_1\Vert\Pb_0)=\int\log\frac{d\Pb_1}{d\Pb_0}\,d\Pb_1\in[0,\infty]$ if $\Pb_1$ is absolutely continuous with respect to $\Pb_0$, and $+\infty$ otherwise; it is measured in nats. We use three standard facts~\cite{cover2006}:
\begin{enumerate}
\item[(F1)] \emph{Chain rule.} For a random vector $(U,V)$, the divergence of its two joint laws is the divergence of the laws of $U$ plus the expected divergence, under $\Pb_1$, of the conditional laws of $V$ given $U$.
\item[(F2)] \emph{Invariance.} If $f:\R^n\to\R^n$ is a bijection such that $f$ and $f^{-1}$ are Borel measurable, then mapping both laws through $f$ does not change their divergence.
\item[(F3)] \emph{Gaussian laws.} For $\mu_1,\mu_0\in\R^n$ and $C\in\Sym^n_{++}$, $\KL\big(\N(\mu_1,C)\Vert\N(\mu_0,C)\big)=\frac12\|\mu_1-\mu_0\|^2_{C^{-1}}$.
\end{enumerate}

\emph{Units.} Time is in seconds (s), frequency in hertz (Hz) and angle in radians (rad); a clock drift is in seconds per second (s/s), a dimensionless ratio. Table~\ref{tab:symbols} lists the main symbols with their domains, their units and the values used in Section~\ref{sec:results}.

\begin{table*}[!t]
\caption{Main symbols, their domains and units, and the values used in Section~\ref{sec:results}. A dash in the unit column marks a dimensionless or component-dependent quantity.}\label{tab:symbols}
\centering\footnotesize
\setlength{\tabcolsep}{4pt}
\renewcommand{\arraystretch}{1.08}
\begin{tabular}{@{}L{2.75cm}L{3.3cm}L{1.55cm}L{6.4cm}L{2.95cm}@{}}
\toprule
Symbol & Domain & Unit & Meaning & Value\\
\midrule
\multicolumn{5}{@{}l}{\emph{Time and indices}}\\
$i$ & $\{r,g\}$ & -- & side of the breaker: utility ($r$) or island ($g$) & --\\
$k$;\ $\Kw_N$ & $\Z$;\ $\{0,\dots,N-1\}$ & -- & frame index; frames of one audit window & --\\
$T_s$ & $(0,\infty)$ & s & frame length & $0.02$\\
$N$;\ $T=NT_s$ & $\Nat$;\ $(0,\infty)$ & --;\ s & audit period in frames; in seconds & to be chosen\\
$L$ & $\Z_{\ge0}$ & -- & number of frames recorded before the window & any\\
\midrule
\multicolumn{5}{@{}l}{\emph{Island and controller}}\\
$f_0$;\ $\omega_0=2\pi f_0$ & $(0,\infty)$ & Hz;\ rad/s & nominal frequency & $50$;\ $314.16$\\
$\theta_{i,k}$;\ $\nu_{i,k}$;\ $x_{i,k}$ & $\R$;\ $\R$;\ $\R^2$ & rad;\ Hz;\ -- & phase; frequency deviation; state $x_{i,k}=(\theta_{i,k},\nu_{i,k})^\top$ of side $i$ & --\\
$w_k$;\ $d_k$;\ $u_k$ & $\R$ & Hz & grid-frequency step; load disturbance; frequency command & --\\
$q_w$;\ $q_d$ & $(0,\infty)$ & Hz$^2$ & variances of $w_k$ and $d_k$ & $10^{-8}$;\ $10^{-6}$\\
$T_g$;\ $\rho$;\ $\beta$ & $(0,\infty)$;\ $(0,1)$;\ $(0,T_s)$ & s;\ --;\ s & governor time constant; $e^{-T_s/T_g}$; $T_g(1-\rho)$ & $0.5$;\ $0.9608$;\ $0.0196$\\
$K_\theta$;\ $K_P$;\ $K_I$ & $\R$ & Hz/rad;\ --;\ 1/s & gains of the PI controller & stabilizing\\
$\varphi_k$;\ $\varphi^{\max}$ & $\R$;\ $(0,\pi)$ & rad & true phase difference across the breaker; reconnection window & --;\ $0.1745$ ($10^\circ$)\\
\midrule
\multicolumn{5}{@{}l}{\emph{Clocks (attacker)}}\\
$\tau_{i,k}$;\ $\gamma_{i,k}$ & $\R$;\ $[-\gb,\gb]$ & s;\ s/s & clock bias of PMU $i$; drift $(\tau_{i,k+1}-\tau_{i,k})/T_s$ & --\\
$\gb$;\ $\Delta=\gb T_s$ & $(0,\infty)$ & s/s;\ s & rate limit of the drift; largest change of bias per frame & $2\times10^{-5}$;\ $4\times10^{-7}$\\
$\tc=\varphi^{\max}/\omega_0$ & $(0,1/(2f_0))$ & s & damaging bias & $5.56\times10^{-4}$\\
$\tau$;\ $\A_N$ & $\A_N\subset\R^{2N}$ & s & clock profile; admissible profiles~\eqref{eq:AN} & --\\
\midrule
\multicolumn{5}{@{}l}{\emph{Reports and monitors (defender)}}\\
$\pth_{i,k}$;\ $\pnu_{i,k}$;\ $y_{i,k}$ & $\R$;\ $\R$;\ $\R^2$ & rad;\ Hz;\ -- & reported phase; reported frequency; $y_{i,k}=(\pth_{i,k},\pnu_{i,k})^\top$ & --\\
$v_{i,k}$;\ $v'_{i,k}$ & $\R$ & rad;\ Hz & PMU noises, with standard deviations $\sigma_\theta,\sigma_\nu\in(0,\infty)$ & $\sigma_\theta=10^{-3}$;\ $\sigma_\nu=10^{-3}$\\
$o_{i,k}$;\ $n_{i,k}$ & $\R$ & s & holdover reading; its noise, with standard deviation $\sigma_o\in(0,\infty)$ & $\sigma_o=2\times10^{-8}$\\
$\delta_i$;\ $\sigma_i$ & $\R$;\ $(0,\infty)$ & s/s & frequency offset of oscillator $i$; its standard deviation & TCXO $3.236\times10^{-7}$, OCXO $3.236\times10^{-9}$, Rb $3.236\times10^{-11}$\\
$\tref_k$;\ $e_k$ & $\R$ & rad & trusted utility-phase reference; its noise, with standard deviation $\sigma_e\in(0,\infty)$ & $\sigma_e=10^{-3}$\\
$\Jset$;\ $J$ & $\subseteq\{r,g,e\}$;\ $\{0,1,2,3\}$ & -- & installed monitors: holdover checks at PMU $r$, $g$ and reference $e$; their number $|\Jset|$ & --\\
$\lambda$;\ $\Dset$ & $(0,\infty)^J$;\ $\Dset\subset(0,\infty)^J$ & (s/s)$^{-2}$, rad$^{-2}$ & design, $\lambda_j=\sigma_j^{-2}$ for $j\in\Jset$, in (s/s)$^{-2}$ for $j=r,g$ and in rad$^{-2}$ for $j=e$; affordable designs & --\\
\midrule
\multicolumn{5}{@{}l}{\emph{Record and evidence}}\\
$\Yp$;\ $Y$ & $\R^{(4+J_e)L}$;\ $\R^m$ & -- & record before frame $0$; record of frames $0,\dots,N-1$, with $m=N(4+J)$ & --\\
$\eta$;\ $\varepsilon$ & $\R^p$;\ $\R^m$ & -- & nuisance, with $p=2N+2+J_o$; measurement noise of the window & --\\
$P$;\ $R$;\ $\Psi$;\ $M$ & $\Sym^p_{++}$;\ $\Sym^m_{++}$;\ $\Sym^m_{++}$;\ $\Sym^{2N}_{++}$ & -- & covariance of $\eta$; of $\varepsilon$; $\Psi=R+FPF^\top$; $M=G^\top\Psi^{-1}G$ & --\\
$P_i$;\ $s_i$ & $\Sym^2_{++}$;\ $(0,\infty)$ & --;\ Hz & steady-state prediction covariance of side $i$; frequency uncertainty & $s_r=2.4\times10^{-4}$, $s_g=1.0\times10^{-4}$\\
$\Ev_\lambda(\tau)$;\ $\Es_\lambda(N)$ & $[0,\infty)$;\ $[0,\infty]$ & nats & evidence of $\tau$; least evidence of a successful attack & --\\
$h$;\ $\alpha$;\ $B$ & $(0,\infty)$;\ $(0,\frac12)$;\ $(1,\infty)$ & nats;\ --;\ -- & evidence threshold; false-alarm probability; detector threshold, a likelihood ratio (on average at most one false alarm per $B$ frames) & $h_Z=4.77$, $h_B=15.3$;\ $10^{-3}$;\ $4.3\times10^6$\\
$T^\star_h(\lambda)$ & $[0,\infty]$ & s & safe audit period & Table~\ref{tab:audit}\\
\bottomrule
\end{tabular}
\end{table*}

\section{The Inspection Game}\label{sec:game}

We first state the game in full. The physics that generates its ingredients (the clocks, the monitors and the defender's record) is derived in Section~\ref{sec:model}; here only the types of these objects matter. Time is divided into frames $k\in\Z$ of length $T_s\in(0,\infty)$, frame $k$ starting at $t_k=kT_s$. Two PMUs, one on the utility side ($i=r$) and one on the island side ($i=g$) of an open breaker, report a phasor once per frame. Their clocks are audited every $N\in\Nat$ frames, and we consider the window from the audit at frame $0$ to the next audit at frame $N$.

\begin{definition}[Inspection game]\label{def:game}
Fix the audit period $N\in\Nat$ and the model of Section~\ref{sec:model}. The \emph{inspection game} $\Gamma_N$ is the two-player zero-sum game with the following elements.
\begin{enumerate}
\item \emph{Players.} The defender $D$, the inspector, operates the island, its controller, its two PMUs and its monitors. The attacker $A$, the inspectee, controls the GNSS signals that the two PMUs receive.
\item \emph{Defender's action.} A \emph{design} $\lambda=(\lambda_j)_{j\in\Jset}\in\Dset$. Here $\Jset\subseteq\{r,g,e\}$ is the set of installed monitors: $j=r$ and $j=g$ are holdover checks at the utility and island PMUs, and $j=e$ is a trusted phase reference (Section~\ref{sec:monitors}). $J=|\Jset|\in\{0,1,2,3\}$ is their number, $\lambda_j\in(0,\infty)$ is the precision of monitor $j$ (the inverse variance of the oscillator-rate offset for a holdover check, or of the measurement noise for the reference), and the set of affordable designs $\Dset\subset(0,\infty)^J$ is nonempty, convex and compact.
\item \emph{Attacker's action.} A \emph{clock profile}
\[
\tau=(\tau_{r,1},\dots,\tau_{r,N},\tau_{g,1},\dots,\tau_{g,N})\in\A_N,
\]
where $\tau_{i,k}\in\R$ is the bias of the clock of PMU $i$ at frame $k$, and
\begin{equation}\label{eq:AN}
\A_N=\big\{\tau\in\R^{2N}:\ \tau\ \text{satisfies (R) and (S)}\big\},
\end{equation}
with the convention $\tau_{i,0}=0$ and the two conditions
\begin{enumerate}
\item[(R)] \emph{rate limit:} $|\tau_{i,k+1}-\tau_{i,k}|\le\Delta$ for all $i\in\{r,g\}$ and $k\in\Kw_N$;
\item[(S)] \emph{success:} $\tau_{r,N}-\tau_{g,N}=\tc$, where $\tau_{i,N}$ is the bias just before the audit at frame $N$ resets it.
\end{enumerate}
The constants $\Delta\in(0,\infty)$, the largest change of bias per frame, and $\tc\in(0,\infty)$, the damaging bias, are derived in Sections~\ref{sec:clock}--\ref{sec:damage}.
\item \emph{Order and information.} (i)~$D$ chooses $\lambda$ and announces it. (ii)~$A$, who knows the model, the controller, $N$ and $\lambda$, chooses $\tau$. The attack is open loop: $A$ observes nothing afterwards. (iii)~Nature draws the initial states and the noises, and the model generates the record $\mathcal Y=(\Yp,Y)$. $D$ observes $\mathcal Y$ but not $\tau$. The record ends at frame $N-1$: the audit at frame $N$, which reveals the bias directly, is not part of $\mathcal Y$.
\item \emph{Payoff.} $A$ pays $D$ the \emph{evidence}
\[
\Ev_\lambda(\tau)=\KL\big(\Pb^\tau_\lambda\,\Vert\,\Pb^0_\lambda\big)\in[0,\infty],
\]
where $\Pb^\tau_\lambda$ is the law of $\mathcal Y$ under design $\lambda$ and profile $\tau$, and $\Pb^0_\lambda$ its law without attack ($\tau=0$).
\end{enumerate}
\end{definition}

\begin{lemma}[Attacker's action set]\label{lem:AN}
For every $N\in\Nat$, $\A_N$ is a compact convex polyhedron, and it is nonempty if and only if $2N\Delta\ge\tc$.
\end{lemma}

\begin{proof}
$\A_N$ is the intersection of finitely many closed half-spaces (R) with a hyperplane (S), hence a closed convex polyhedron. By (R) and $\tau_{i,0}=0$, $|\tau_{i,k}|\le k\Delta\le N\Delta$, so $\A_N$ is bounded and $\tau_{r,N}-\tau_{g,N}\le2N\Delta$. Conversely, if $2N\Delta\ge\tc$, the opposite ramps $\tau_{r,k}=k\tc/(2N)$ and $\tau_{g,k}=-k\tc/(2N)$ change by $\tc/(2N)\le\Delta$ per frame and satisfy (S).
\end{proof}

\emph{Values and saddle points.} For a design $\lambda$, the attacker's \emph{best-response value}, the least evidence of a successful attack, is
\[
\Es_\lambda(N)=\inf_{\tau\in\A_N}\Ev_\lambda(\tau)\in[0,\infty],
\]
with $\Es_\lambda(N)=+\infty$ if $\A_N=\emptyset$. The value of $\Gamma_N$ when $D$ leads, as in item~4, is $\sup_{\lambda\in\Dset}\Es_\lambda(N)$. A pair $(\lambda^\star,\tau^\star)\in\Dset\times\A_N$ is a \emph{saddle point} if $\Ev_\lambda(\tau^\star)\le\Ev_{\lambda^\star}(\tau^\star)\le\Ev_{\lambda^\star}(\tau)$ for all $\lambda\in\Dset$ and $\tau\in\A_N$: neither player gains by deviating alone, and the order of moves is immaterial.

\begin{definition}[Safe audit period]\label{def:safe}
Let $h\in(0,\infty)$ be an evidence threshold and $\lambda\in(0,\infty)^J$ a design. Audits every $N\in\Nat$ frames are \emph{$h$-safe} for $\lambda$ if $\Es_\lambda(N')\ge h$ for every $N'\in\{1,\dots,N\}$. The \emph{safe audit period} is
\[
T^\star_h(\lambda)=T_s\cdot\sup\mathcal S_h(\lambda)\in[0,\infty],
\]
where $\mathcal S_h(\lambda)\subseteq\Nat$ is the set of $N$ for which audits every $N$ frames are $h$-safe for $\lambda$. The defender's \emph{design problem} is to maximize $T^\star_h(\lambda)$ over $\lambda\in\Dset$.
\end{definition}

Audits are thus safe if every successful attack reveals at least the evidence $h$. Requiring this for every shorter period as well lets the defender audit earlier than scheduled without losing safety. Since $\Es_\lambda(N')=+\infty$ whenever $2N'\Delta<\tc$, every period shorter than $\tc/(2\gb)$ seconds is safe: the attacker cannot reach $\tc$ in time.

\begin{figure}[t]
\centering
\begin{tikzpicture}[paper, x=1cm, y=1cm,
  move/.style={draw=cInk!55, line width=0.4pt, rounded corners=2pt, fill=white,
               inner sep=2.6pt, align=center, font=\footnotesize},
  num/.style={circle, fill=cInk, text=white, inner sep=0.6pt, font=\scriptsize\bfseries}]
\fill[cAtk!7] (0,0) -- (4.875,1.95) -- (6,1.5) -- (1.125,-0.45) -- cycle;
\draw[cAtk!40, line width=0.4pt] (0,0) -- (4.875,1.95) -- (6,1.5) -- (1.125,-0.45) -- cycle;
\draw[cMuted!60, line width=0.4pt] (-1.75,0) -- (6.35,0) node[right, text=cInk] {$0$};
\draw[cMuted!60, line width=0.4pt, densely dashed] (-0.15,1.5) -- (6.35,1.5) node[right, text=cInk] {$\tc$};
\foreach \x/\l in {0/{audit, $k=0$}, 6/{audit, $k=N$}}{
  \draw[cInk, line width=0.9pt] (\x,-0.72) -- (\x,2.2);
  \node[above, inner sep=1.5pt] at (\x,2.2) {\l};}
\draw[cAtk, line width=1.1pt] (-1.75,0) -- (0,0) -- (6,1.5);
\node[text=cInk, anchor=south, inner sep=1.5pt] at (-0.9,0.06) {$\tau_{r,k}-\tau_{g,k}$};
\draw[cAtk, line width=1.1pt, dash pattern=on 3pt off 1.8pt] (0,0) -- (2.25,0) -- (6,1.5);
\node[text=cInk, rotate=14, anchor=south, inner sep=1.5pt] at (3.3,0.83) {ramp};
\node[text=cInk, align=left, anchor=north west, inner sep=1pt] at (4.25,0.62) {wait, then\\[-1pt]sprint};
\node[text=cInk, rotate=21.8, anchor=south, inner sep=1.5pt] at (1.25,0.5) {slope $2\Delta$};
\node[text=cAtk, font=\scriptsize] at (1.12,-0.2) {feasible};
\draw[->, cInk, line width=0.5pt] (-1.75,-0.72) -- (6.6,-0.72) node[right, inner sep=1.5pt] {$k$};
\draw[decorate, decoration={brace, mirror, amplitude=3.5pt}, cInk!80, line width=0.4pt]
  (-1.7,-0.86) -- (-0.07,-0.86) node[midway, below=4pt] {past record $\Yp$};
\draw[decorate, decoration={brace, mirror, amplitude=3.5pt}, cInk!80, line width=0.4pt]
  (0.07,-0.86) -- (5.93,-0.86) node[midway, below=4pt] {window record $Y$, frames $0,\dots,N-1$};
\node[move] (s1) at (-0.95,3.15) {$D$ announces\\[-1pt] its design $\lambda$};
\node[move] (s2) at (1.95,3.15) {$A$ commits to a\\[-1pt] profile $\tau\in\A_N$};
\node[move] (s3) at (5.05,3.15) {$D$ records $\mathcal Y$;\\[-1pt] payoff $\Ev_\lambda(\tau)$};
\foreach \s/\n in {s1/1,s2/2,s3/3} \node[num] at (\s.north west) {\n};
\draw[sig] (s1) -- (s2); \draw[sig] (s2) -- (s3);
\end{tikzpicture}
\caption{One round of the inspection game $\Gamma_N$. (1)~The defender announces its design $\lambda$. (2)~Right after the audit at frame $0$, the attacker commits to a clock profile $\tau\in\A_N$. The shaded region contains the differential bias $\tau_{r,k}-\tau_{g,k}$ of every admissible profile: by (R) it changes by at most $2\Delta$ per frame, and by (S) it equals $\tc$ just before the next audit, which then resets it. A ramp and a wait-then-sprint profile, which rides the boundary of the region, are shown. (3)~The defender records $\mathcal Y=(\Yp,Y)$, and the payoff is the evidence $\Ev_\lambda(\tau)$, which measures how distinguishable $\mathcal Y$ is from a record without attack.}
\label{fig:game}
\end{figure}
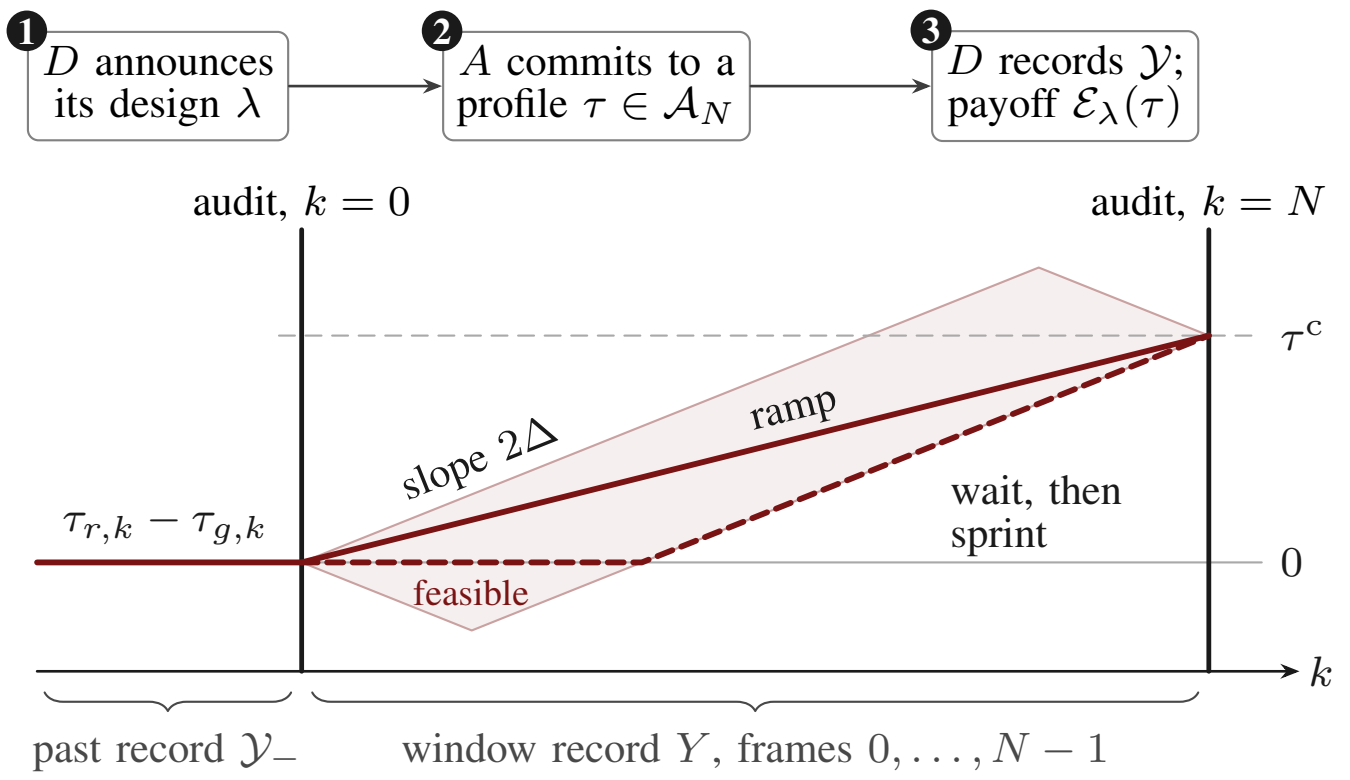

\subsection{Remarks on the game}

\emph{Why the evidence is the payoff.} For each specified attack, the Neyman--Pearson lemma fixes the best achievable probability of detection, and by Proposition~\ref{prop:detect} below it is an increasing function of the evidence; we use the evidence as the design payoff. The design $\lambda$ shapes it, and the attacker, who wants to succeed without being detected, minimizes it. With $h=h_Z=\frac12z_{1-\alpha}^2$, safety is necessary for any test with false-alarm probability $\alpha$ to detect every successful attack with probability at least $\frac12$, and sufficient for the test matched to each attack (Proposition~\ref{prop:detect}). A target probability $p\in[\frac12,1)$ instead requires $h_p=\frac12(z_{1-\alpha}+z_p)^2$. The defender does not observe $\tau$, however, so the matched test is a benchmark rather than an implementable detector. One detector for every admissible profile is a composite test, whose calibration we leave to future work: a safe audit period certifies the evidence of every successful attack, not the power of such a detector.

\emph{Modeling choices.} The controller, the audit schedule and the threshold $h$ are fixed. The attack is open loop: it is planned in advance, as by a spoofer that drags the clock along a preplanned path. Adaptive attacks are left to future work.

\emph{Roadmap.} Section~\ref{sec:model} builds the model that defines $\A_N$ and $\Pb^\tau_\lambda$. Section~\ref{sec:evidence} computes the payoff $\Ev_\lambda(\tau)$. Section~\ref{sec:attacker} solves the attacker's problem $\min_{\tau\in\A_N}\Ev_\lambda(\tau)$, and Section~\ref{sec:design} the defender's problem $\max_{\lambda\in\Dset}\Es_\lambda(N)$. Section~\ref{sec:results} certifies the safe audit periods of a 50\,Hz island.

\section{The Model Behind the Game}\label{sec:model}

\begin{figure}[t]
\centering
\begin{tikzpicture}[paper, x=1cm, y=1cm,
  blk/.style={draw=cInk!70, line width=0.5pt, rounded corners=1.8pt, align=center,
              inner sep=2.4pt, minimum height=8.5mm, fill=white},
  plant/.style={blk, fill=cInk!6},
  pmuR/.style={blk, draw=cUtil, fill=cUtil!9, line width=0.6pt},
  pmuG/.style={blk, draw=cIsl, fill=cIsl!11, line width=0.6pt},
  spoof/.style={blk, draw=cAtk, fill=cAtk!7, line width=0.6pt},
  rec/.style={blk, draw=cInk, line width=0.6pt, minimum width=4.3cm},
  refl/.style={->, draw=cRef, line width=0.7pt, dash pattern=on 2.6pt off 1.6pt}]
\node[rec] (det) at (4.25,3.25) {defender's record $\mathcal Y=(y,\,o,\,\tref)$\\[1.5pt]alarm test; design $\lambda_j=\sigma_j^{-2}$, $j\in\Jset$};
\node[pmuR] (pr) at (1.25,1.85) {PMU $r$\\[-1pt]clock $t+\tau_r$\\[-1pt]+ holdover osc.\ $\sigma_r$};
\node[spoof] (sp) at (4.25,1.85) {GNSS spoofer\\[-1pt]drift $|\gamma_{i,k}|\le\bar\gamma$};
\node[pmuG] (pg) at (7.25,1.85) {PMU $g$\\[-1pt]clock $t+\tau_g$\\[-1pt]+ holdover osc.\ $\sigma_g$};
\node[blk] (ctrl) at (4.25,0.6) {PI controller};
\node[plant] (grid) at (1.25,-0.6) {utility grid\\[-1pt]$\theta_r,\ \nu_r$};
\node[plant] (isl) at (7.25,-0.6) {island generator\\[-1pt]$\theta_g,\ \nu_g$};
\coordinate (b1) at (3.9,-0.6); \coordinate (b2) at (4.6,-0.6);
\draw[cInk!75, line width=1.3pt] (grid.east) -- (b1) (b2) -- (isl.west);
\draw[cInk!75, line width=1.3pt] (b1) -- ++(25:0.68);
\fill[cInk] (b1) circle (1.4pt) (b2) circle (1.4pt);
\node[below=3pt, text=cInk] at ($(b1)!0.5!(b2)$) {open breaker};
\draw[sig] ($(isl.south)+(0,-0.45)$) node[below, inner sep=1pt] {load $d_k$} -- (isl.south);
\draw[sig] (grid.north) -- (pr.south);
\draw[sig] (isl.north) -- (pg.south);
\draw[atk] (sp.west) -- node[above, text=cAtk] {$\tau_r$} (pr.east);
\draw[atk] (sp.east) -- node[above, text=cAtk] {$\tau_g$} (pg.west);
\draw[sig] ([xshift=6mm]pr.south) |- node[pos=0.78, above, inner sep=1.5pt] {$y_r$} (ctrl.west);
\draw[sig] ([xshift=-6mm]pg.south) |- node[pos=0.78, above, inner sep=1.5pt] {$y_g$} ([yshift=1.2mm]ctrl.east);
\draw[sig] ([yshift=-1.4mm]ctrl.east) -| node[pos=0.22, below, inner sep=1.5pt] {$u_k$} ([xshift=-6mm]isl.north);
\draw[sig] ([xshift=-2mm]pr.north) |- node[pos=0.25, left, inner sep=1.5pt] {$y_r$} ([yshift=-1.5mm]det.west);
\draw[sig] ([xshift=3mm]pr.north) -- node[right, inner sep=1.5pt] {$o_r$} ([xshift=3mm]pr.north |- det.south);
\draw[sig] ([xshift=2mm]pg.north) |- node[pos=0.25, right, inner sep=1.5pt] {$y_g$} ([yshift=-1.5mm]det.east);
\draw[sig] ([xshift=-3mm]pg.north) -- node[left, inner sep=1.5pt] {$o_g$} ([xshift=-3mm]pg.north |- det.south);
\coordinate (xl) at ([xshift=-2.4mm]pr.west);
\draw[refl] (grid.west) -- (grid.west -| xl) |- node[pos=0.7, above, inner sep=1.5pt, text=cRef] {$\tref_k$,\ $\sigma_e$} ([yshift=1.8mm]det.west);
\end{tikzpicture}
\caption{A synchronous island under a GNSS timing attack. The PI controller holds the island phase equal to the utility phase, as reported by two GNSS-timed PMUs, while the spoofer drags their clocks. The defender records the PMU reports $y$ and the readings of its monitors $j\in\Jset\subseteq\{r,g,e\}$: the holdover comparisons $o$ at PMU $r$ and $g$ ($j=r,g$, oscillator spreads $\sigma_r,\sigma_g$) and, optionally, a trusted utility-phase reference $\tref$ ($j=e$, noise $\sigma_e$; magenta, dashed). Its design sets the precisions $\lambda_j=\sigma_j^{-2}$. Fig.~\ref{fig:results} uses the same colors for the utility side, the island side and the trusted reference.}
\label{fig:system}
\end{figure}
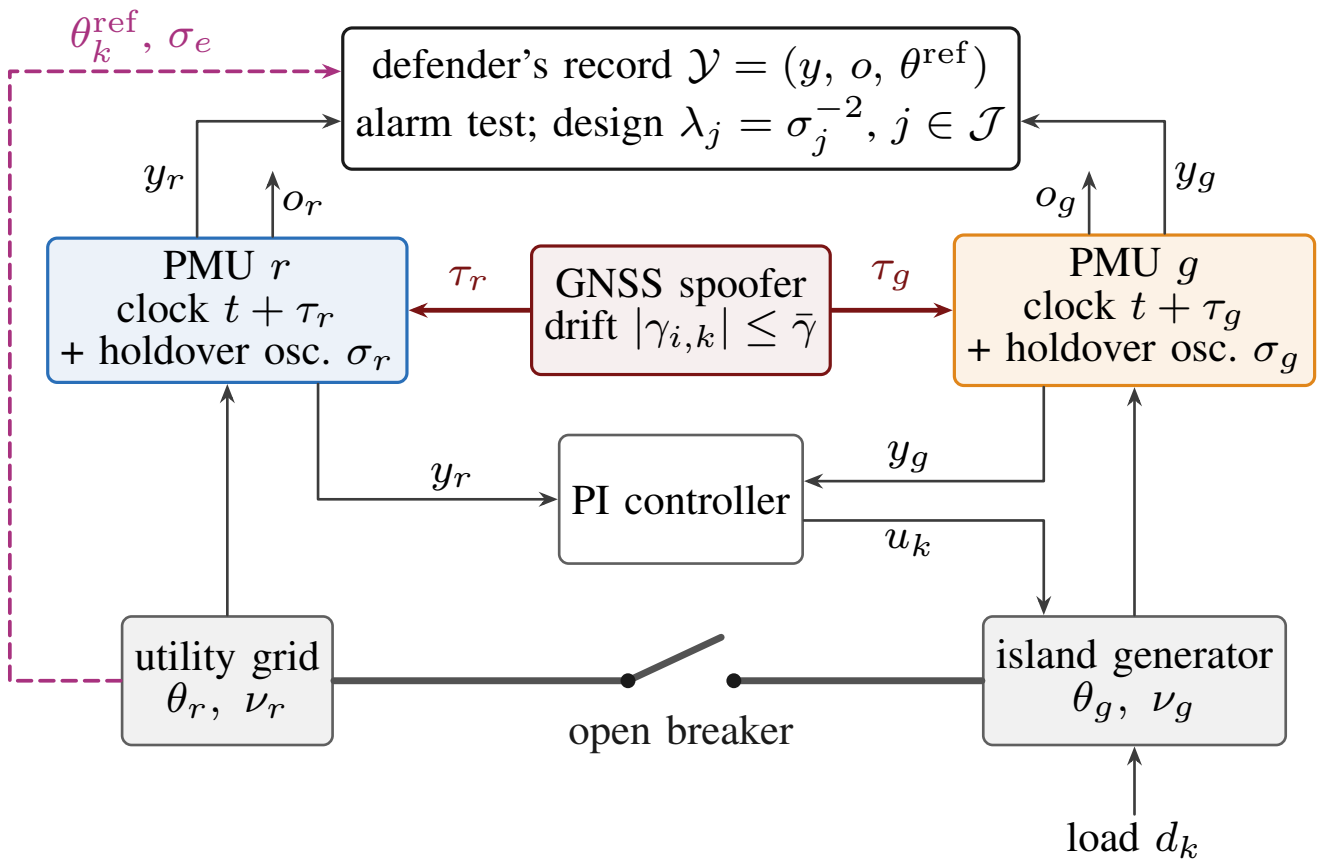

Fig.~\ref{fig:system} shows the setting. We describe the island and its controller, then the clocks that the attacker drags, and finally the record that the defender keeps.

\subsection{The island and its controller}\label{sec:plant}

The nominal frequency is $f_0\in(0,\infty)$ and $\omega_0=2\pi f_0$. Angles are measured against a reference that rotates at $f_0$, so a bus running exactly at $f_0$ has a constant angle; angles are not wrapped. Side $i\in\{r,g\}$ has, at frame $k\in\Z$, phase $\theta_{i,k}\in\R$ (rad), frequency deviation $\nu_{i,k}\in\R$ (Hz) from $f_0$, and state $x_{i,k}=(\theta_{i,k},\nu_{i,k})^\top\in\R^2$.

\emph{Utility grid.} The grid is much larger than the island, and its frequency wanders slowly as a random walk with steps $w_k\in\R$ (Hz), the usual model in Kalman-filter frequency trackers~\cite{girgis1990,ferrero2020}:
\begin{equation}\label{eq:util}
\theta_{r,k+1}=\theta_{r,k}+2\pi T_s\nu_{r,k},\qquad \nu_{r,k+1}=\nu_{r,k}+w_k .
\end{equation}

\emph{Island.} We lump the prime mover, its speed governor and the rotor inertia into a first-order speed loop, the simplest of the low-order frequency-response models~\cite[Ch.~11]{kundur1994},~\cite{anderson1990}. The island generator thus follows its frequency command $u\in\R$ (Hz) with time constant $T_g\in(0,\infty)$, disturbed by the load. We express the load disturbance $d\in\R$ as the frequency deviation that it would cause in steady state, so that $\dot\theta_g=2\pi\nu_g$ and $T_g\dot\nu_g=-\nu_g+u+d$. Holding $u=u_k$ and $d=d_k$ constant during frame $k$ and integrating over the frame gives the exact sampled model
\begin{equation}\label{eq:isl}
\begin{aligned}
\theta_{g,k+1}&=\theta_{g,k}+2\pi\beta\,\nu_{g,k}+2\pi(T_s-\beta)(u_k+d_k),\\
\nu_{g,k+1}&=\rho\,\nu_{g,k}+(1-\rho)(u_k+d_k),
\end{aligned}
\end{equation}
where $\rho=e^{-T_s/T_g}\in(0,1)$ is the fraction of a frequency deviation that survives one frame, and $\beta=T_g(1-\rho)$. Since $0<1-e^{-x}<x$ for $x>0$, $\beta\in(0,T_s)$. In matrix form, with
\[
\begin{aligned}
A_r&=\begin{pmatrix}1&2\pi T_s\\0&1\end{pmatrix},& b_r&=\begin{pmatrix}0\\1\end{pmatrix},\\
A_g&=\begin{pmatrix}1&2\pi\beta\\0&\rho\end{pmatrix},& b_g&=\begin{pmatrix}2\pi(T_s-\beta)\\1-\rho\end{pmatrix},
\end{aligned}
\]
\eqref{eq:util}--\eqref{eq:isl} read $x_{r,k+1}=A_rx_{r,k}+b_rw_k$ and $x_{g,k+1}=A_gx_{g,k}+b_g(u_k+d_k)$. Nothing couples the two sides while the breaker is open.

\emph{Controller.} The controller acts on what the PMUs \emph{report}, the phase $\pth_{i,k}$ and the frequency $\pnu_{i,k}$ of Section~\ref{sec:pmu}, through the proportional--integral (PI) law:
\begin{equation}\label{eq:ctrl}
\begin{aligned}
\zeta_k&=(\pnu_{r,k}-\pnu_{g,k})+K_\theta(\pth_{r,k}-\pth_{g,k}),\\
u_k&=K_P\zeta_k+K_I\xi_k,\qquad \xi_{k+1}=\xi_k+T_s\zeta_k ,
\end{aligned}
\end{equation}
with error $\zeta_k\in\R$ (Hz), integrator state $\xi_k\in\R$ (Hz\,s) and gains $K_\theta,K_P,K_I\in\R$ for which the island part of the closed loop is stable: with the utility state, the load, the noises and the attack as exogenous inputs, the state $(\theta_{g,k},\nu_{g,k},\xi_k)$ obeys a linear recursion whose matrix has all its eigenvalues in the open unit disk. (The utility side, a random walk, is not stabilized by anything; the loop only makes the island follow it.) Since frequency is the rate of phase divided by $2\pi$, the error $\zeta_k$ is a proportional--derivative term in the reported phase difference, which drives a PI loop on the frequency command; laboratory implementations of synchronous islanding likewise add a phase-difference term to the speed reference of the governor~\cite{best2007,best2011}. Our results hold for a much larger class. A controller is \emph{admissible} if, for every frame $k$, $u_k=\pi_k(\text{record up to frame }k)$ for a Borel-measurable map $\pi_k$; the PI law~\eqref{eq:ctrl} is admissible. What matters below is only that $u_k$ is computed from the data, and that the integral action of~\eqref{eq:ctrl} drives the reported phase difference to zero.

\subsection{The clocks: what the attacker controls}\label{sec:clock}

PMU $i$ takes its time from a GNSS receiver whose clock reads $t+\tau_i$ at true time $t$. The \emph{bias} $\tau_{i,k}\in\R$ (s) at frame $k$ is positive when the clock is ahead. A spoofer that has captured the receiver can drag the clock, but only slowly, because a fast change breaks the receiver's lock or trips its clock-consistency checks~\cite{humphreys2008,psiaki2016}. Against the timing receiver of a PMU, a drag of 1000\,m/s in range-equivalent units, that is $3.3$\,\textmu s of bias per second, has been demonstrated~\cite{shepard2012}, and a power-grid timing receiver lost lock above 400\,m/s ($1.3$\,\textmu s/s)~\cite{shepard2011}. The attacker therefore controls the \emph{drift} $\gamma_{i,k}\in\R$ (s/s), the rate of change of the bias, up to a rate limit $\gb\in(0,\infty)$:
\begin{equation}\label{eq:clock}
\tau_{i,k+1}=\tau_{i,k}+T_s\gamma_{i,k},\qquad |\gamma_{i,k}|\le\gb .
\end{equation}
For example, a drift of $10^{-5}$ adds 10\,\textmu s of bias per second. The bias changes by at most $\Delta=\gb T_s$ per frame, which is condition (R) of Definition~\ref{def:game}. At frames $0$ and $N$, both clocks are \emph{audited} against an independent time source, which resets the biases and reveals any bias that matters. We write $\tau_{i,N}$ for the bias just before the audit at frame $N$ resets it, and the defender's record ends at frame $N-1$; the audit itself is not part of the evidence, since an exact audit would reveal any bias directly. Closing the breaker at an earlier frame $N'<N$ is covered by Definition~\ref{def:safe}, which requires safety for every $N'\le N$. The attacker knows the schedule and starts right after an audit, so $\tau_{i,k}=0$ for $k\le0$. It fixes its plan, the profile $\tau\in\R^{2N}$, in advance; the drifts $\gamma_{i,k}=(\tau_{i,k+1}-\tau_{i,k})/T_s$ are then linear in $\tau$.

\subsection{What the PMUs report}\label{sec:pmu}

A PMU reports the angle of the bus voltage relative to a cosine at $f_0$ that it generates from its own clock~\cite{c37118}. If its clock is ahead by $\tau_i$, its reference cosine is advanced by $\omega_0\tau_i$, so the voltage appears to lag by $\omega_0\tau_i$. Frequency is the rate of change of angle divided by $2\pi$, so the reported frequency is lower by $f_0$ times the drift. The reports at frame $k$ are therefore
\begin{equation}\label{eq:pmu}
\pth_{i,k}=\theta_{i,k}-\omega_0\tau_{i,k}+v_{i,k},\qquad
\pnu_{i,k}=\nu_{i,k}-f_0\gamma_{i,k}+v'_{i,k},
\end{equation}
with measurement noises $v_{i,k}\in\R$ (rad) and $v'_{i,k}\in\R$ (Hz), and we write $y_{i,k}=(\pth_{i,k},\pnu_{i,k})^\top\in\R^2$. The attack thus adds to the reports a \emph{signature} that is fixed by the clock profile. Equation~\eqref{eq:pmu} omits terms of relative order $|\nu_{i,k}|/f_0$ and $\gb$, which arise because the waveform is sampled at the wrong instant by a clock that runs at the wrong rate; they are below $2\times10^{-3}$ for frequency deviations up to $0.1$\,Hz.

\subsection{Why the attack causes damage}\label{sec:damage}

Let $\varphi_k=\theta_{r,k}-\theta_{g,k}\in\R$ be the true phase difference across the breaker. By~\eqref{eq:pmu}, the controller sees $\varphi_k-\omega_0(\tau_{r,k}-\tau_{g,k})$ plus noise. If the differential bias $\tau_r-\tau_g$ is held constant, the integral action drives what the controller sees to zero on average, so the true difference settles at $\omega_0(\tau_r-\tau_g)$ while the controller reports perfect synchronism. Retuning the gains does not help: a slowly drifting differential bias looks to the loop like a change of grid frequency, which it is designed to follow. Closing the breaker is safe only if $|\varphi|$ is below the reconnection window $\varphi^{\max}\in(0,\pi)$. We therefore call an attack \emph{successful} if it reaches the \emph{damaging bias}
\begin{equation}\label{eq:target}
\tau_{r,N}-\tau_{g,N}=\tc,\qquad \tc=\varphi^{\max}/\omega_0\in\big(0,\tfrac1{2f_0}\big),
\end{equation}
just before the next audit, which is condition (S) of Definition~\ref{def:game}. For $\varphi^{\max}=10^\circ=0.1745$\,rad at 50\,Hz, $\tc=0.556$\,ms. The sign in~\eqref{eq:target} is without loss of generality.

\emph{From bias to phase.} Condition (S) constrains the differential bias, not the phase itself: it is a proxy for an unsafe reconnection, whose accuracy we now bound. The closed loop is linear, so $\varphi_k$ is the sum of its value without attack, whose law the attack does not change, and of a deterministic response $\varphi^a_k$ to the signature. Let $D_k=\tau_{r,k}-\tau_{g,k}$, and let $\omega_0(1+\varsigma_m)$, $m\in\Z_{\ge0}$, be the value of $\varphi^a$ at frame $k_0+m$ when, from a zero attack response, $D$ steps from $0$ to $1$ at frame $k_0$, so that $\varsigma_m$ is the relative tracking error of a step. The frequency report of frame $k_0-1$ already carries the drift $(D_{k_0}-D_{k_0-1})/T_s$, so the command $u_{k_0-1}$ reacts and $\varphi^a$ responds at frame $k_0$ itself ($m=0$). Superposing steps, with $D_0=0$,
\[
\varphi^a_N=\omega_0D_N+\omega_0\sum_{k=1}^{N}\varsigma_{N-k}\,(D_k-D_{k-1}),
\]
and since $|D_k-D_{k-1}|\le2\Delta$ by (R), every profile that satisfies (R) has
\[
\big|\varphi^a_N-\omega_0D_N\big|\le2\Delta\,\omega_0\,\|\varsigma\|_1,\qquad \|\varsigma\|_1=\textstyle\sum_{m\ge0}|\varsigma_m|.
\]
The sum is finite because the island part of the loop is stable and its integral action drives the reported phase difference, and hence $\varsigma_m$, to zero geometrically. With $T_g=0.5$\,s and the gains $K_\theta=0.2$\,Hz/rad, $K_P=0.5$ and $K_I=0.2$\,s$^{-1}$ of a gentle tuning A, the island part of the sampled loop has spectral radius $0.994$ and $\|\varsigma\|_1=93.6$ frames, so every admissible profile shifts the phase difference at frame $N$ by $10^\circ$ to within $1.35^\circ$. For a faster tuning B ($K_\theta=1$\,Hz/rad, $K_P=2$, $K_I=1$\,s$^{-1}$), $\|\varsigma\|_1=13.7$ frames and the margin is $0.20^\circ$. The worst case needs a drift that reverses in step with the transient of the loop, and specific profiles do far better. Under tuning A, $\varphi^a_N-\omega_0\tc$ is $-5.0\times10^{-9}$ degrees for a differential ramp over 60\,s; $4.5\times10^{-5}$ for one clock that sprints at $\gb$ in the last 27.8\,s of a 60\,s window; $-1.2\times10^{-4}$ and $-5.3\times10^{-4}$ for differential ramps over 26.08 and 24.28\,s, the safe periods without monitors of Section~\ref{sec:results}; and $0.025$ for the fastest feasible ramp, over 13.9\,s. Under tuning B, all five are below $4\times10^{-5}$ degrees. With noise, the phase difference at reconnection is this deterministic shift plus the same random fluctuation as without attack. To guarantee an attack-induced shift of at least $\varphi^{\max}$ for every profile shape, the attacker needs at most the extra bias $2\Delta\|\varsigma\|_1$; a guarantee for the realized phase would also need a chance constraint on that fluctuation. By Lemma~\ref{lem:AN}, reaching $\tc$ takes at least $\tc/(2\gb)$ seconds if both clocks are dragged in opposite directions, and $\tc/\gb$ seconds with one clock.

\subsection{The monitors and the design}\label{sec:monitors}

The defender records the PMU reports. A \emph{monitoring design} adds channels whose time does not come from GNSS:
\begin{itemize}
\item A \emph{holdover check} at PMU $i\in\{r,g\}$ (monitor $j=i$). A local oscillator, synchronized at the audit, then runs freely, and the PMU records GNSS time minus oscillator time. The oscillator's rate is off by an unknown fraction $\delta_i\in\R$ (s/s): at true time $t$ after the audit it reads $(1-\delta_i)t$, so $\delta_i>0$ when it runs slow. The comparison therefore reads, for $k\in\Kw_N$,
\begin{equation}\label{eq:holdover}
o_{i,k}=\tau_{i,k}+\delta_i\,kT_s+n_{i,k},
\end{equation}
with comparison noise $n_{i,k}\in\R$ (s). Better oscillators have a smaller spread $\sigma_i\in(0,\infty)$ of $\delta_i$: in increasing order of quality, temperature-compensated (TCXO), oven-controlled (OCXO) and rubidium (Rb) oscillators~\cite{lombardi2002}. Timing receivers already use such a comparison: one with an OCXO enters holdover when GNSS and oscillator time differ by more than 1\,\textmu s, which limited a spoofer to a drag of 2\,m/s~\cite{shepard2011}.
\item A \emph{trusted reference} (monitor $j=e$): a measurement of the utility phase timed independently of GNSS, for $k\in\Z$,
\begin{equation}\label{eq:ref}
\tref_k=\theta_{r,k}+e_k,
\end{equation}
with noise $e_k\in\R$ (rad) of standard deviation $\sigma_e\in(0,\infty)$. It can come, for example, from a PMU synchronized over fiber~\cite{derviskadic2020}, provided that the link is secured against delay attacks~\cite{narula2018}.
\end{itemize}
The installed monitors form a set $\Jset\subseteq\{r,g,e\}$, where $j\in\{r,g\}$ stands for the holdover check at PMU $j$ and $j=e$ for the trusted reference. Let $J_o=|\Jset\cap\{r,g\}|\in\{0,1,2\}$ be the number of holdover checks, $J_e=|\Jset\cap\{e\}|\in\{0,1\}$ the number of references, and $J=|\Jset|=J_o+J_e$. Each monitor $j\in\Jset$ has a noise spread $\sigma_j$: the oscillator spread $\sigma_r$ or $\sigma_g$ in~\eqref{eq:holdover}, or the reference noise $\sigma_e$ in~\eqref{eq:ref}. The design lists their precisions,
\[
\lambda_j=\sigma_j^{-2},\qquad j\in\Jset,
\]
in the order $r,g,e$, so that $\lambda\in(0,\infty)^J$ and a better instrument has a larger precision. For example, with both holdover checks and the reference, $\Jset=\{r,g,e\}$ and $\lambda=(\sigma_r^{-2},\sigma_g^{-2},\sigma_e^{-2})$; with the reference alone, $\Jset=\{e\}$ and $\lambda=\sigma_e^{-2}$. The set $\Dset$ of Definition~\ref{def:game} models what the defender can afford. It is a single point for fixed hardware. When precision can be bought, a budget such as $\{\lambda\in[\epsilon,\infty)^J:\sum_{j\in\Jset}c_j\lambda_j\le C\}$ is nonempty, convex and compact, where $c_j\in(0,\infty)$ is the cost of a unit of precision of monitor $j$, $C\in(0,\infty)$ the total budget and $\epsilon\in(0,\infty)$ with $\epsilon\sum_{j\in\Jset}c_j\le C$. The weights $c_j$ make precisions of different units, (s/s)$^{-2}$ for $j=r,g$ and rad$^{-2}$ for $j=e$, commensurable; with two oscillators alone, equal weights are meaningful. Convex increasing costs $c_j(\lambda_j)$ in place of $c_j\lambda_j$ keep $\Dset$ convex.

\subsection{The record and the assumptions}\label{sec:noise}

The defender records the PMU reports and the reference, if installed, from frame $-L$ on, where $L\in\Z_{\ge0}$; the holdover checks are read during the window only, because the oscillators are resynchronized at each audit. The \emph{past record} $\Yp$ collects $y_{r,k}$, $y_{g,k}$ and $\tref_k$ for $-L\le k\le-1$. The \emph{window record} $Y=(Y_0,\dots,Y_{N-1})\in\R^m$ collects, for each $k\in\Kw_N$, the vector $Y_k$ of $y_{r,k}$, $y_{g,k}$, the installed $o_{i,k}$ and, if installed, $\tref_k$; thus $m=N(4+J)$. The commands need not be recorded, since by admissibility they are computed from the record.

\begin{assumption}[Linear model]\label{as:lin}
For all $k\ge-L$, the plant~\eqref{eq:util}--\eqref{eq:isl}, an admissible controller, the clocks~\eqref{eq:clock}, the reports~\eqref{eq:pmu} and the monitors~\eqref{eq:holdover}--\eqref{eq:ref} hold exactly, without saturation, with $\tau_{i,k}=0$ for $k\le0$.
\end{assumption}

Assumption~\ref{as:lin} makes~\eqref{eq:pmu} exact within the model, although Section~\ref{sec:pmu} derived it by neglecting terms of relative order $2\times10^{-3}$. Our analysis is thus exact for the adopted model, not for the physical observation law.

\begin{assumption}[Gaussian noise]\label{as:gauss}
The initial states $x_{r,-L},x_{g,-L}$, the offsets $\delta_r,\delta_g$ and all the samples $w_k$, $d_k$, $v_{i,k}$, $v'_{i,k}$, $n_{i,k}$ and $e_k$, over all indices $i$ and $k$, are mutually independent, so that each sequence is white; moreover, $w_k\sim\N(0,q_w)$, $d_k\sim\N(0,q_d)$, $v_{i,k}\sim\N(0,\sigma_\theta^2)$, $v'_{i,k}\sim\N(0,\sigma_\nu^2)$, $n_{i,k}\sim\N(0,\sigma_o^2)$, $e_k\sim\N(0,\sigma_e^2)$ and $\delta_i\sim\N(0,\sigma_i^2)$, with $q_w,q_d,\sigma_\theta,\sigma_\nu,\sigma_o,\sigma_e,\sigma_i\in(0,\infty)$. These laws depend neither on the attack nor on the controller.
\end{assumption}

White, stationary load fluctuations are the standard model of ambient power-system data~\cite{pierre1997}, and PMU errors have been found close to Gaussian~\cite{brown2016}, although field data also show heavier tails~\cite{wang2018}. Assumption~\ref{as:gauss} makes the evidence a quadratic form; Section~\ref{sec:conclusion} returns to the load model.

Before the window, each side is a linear system observed in Gaussian noise. Side $i$ is measured through $C_ix_{i,k}$ plus noise of covariance $R_i$, where $C_g=I_2$ and $R_g=\mathrm{diag}(\sigma_\theta^2,\sigma_\nu^2)$; $C_r=I_2$ and $R_r=R_g$ without the reference; and $C_r=\big(\begin{smallmatrix}I_2\\ 1\ 0\end{smallmatrix}\big)$ and $R_r=\mathrm{diag}(\sigma_\theta^2,\sigma_\nu^2,\sigma_e^2)$ with it. Its process noise has covariance $Q_r=q_wb_rb_r^\top$ or $Q_g=q_db_gb_g^\top$. For $\Pi\in\Sym^2_+$, let
\[
\begin{aligned}
\Ric_i(\Pi)={}&A_i\Pi A_i^\top+Q_i\\
&-A_i\Pi C_i^\top(C_i\Pi C_i^\top+R_i)^{-1}C_i\Pi A_i^\top
\end{aligned}
\]
be the Riccati map of side $i$, which updates the Kalman filter's prediction covariance over one frame~\cite{anderson1979}.

\begin{lemma}[Steady-state prediction]\label{lem:kalman}
Let $i\in\{r,g\}$.
\begin{enumerate}
\item[(a)] The equation $\Pi=\Ric_i(\Pi)$ has exactly one solution $P_i\in\Sym^2_+$ for which the matrix $A_i-A_iP_iC_i^\top(C_iP_iC_i^\top+R_i)^{-1}C_i$ has all its eigenvalues in the open unit disk, and $P_i\in\Sym^2_{++}$.
\item[(b)] Let~\ref{as:lin}--\ref{as:gauss} hold and $x_{i,-L}\sim\N(\bar x_i,\Pi_i)$ with $\bar x_i\in\R^2$ and $\Pi_i\in\Sym^2_+$. Without attack, given $\Yp$, the prediction errors $x_{i,0}-\hat x_{i,0}$, where $\hat x_{i,0}=\Ex[x_{i,0}\mid\Yp]$, are independent across $i$ and Gaussian with mean $0$ and covariance $\Ric_i^L(\Pi_i)$, the $L$-fold composition. This covariance depends neither on $\Yp$, nor on $\bar x_i$, nor on the controller, and $\Ric_i^L(\Pi_i)\to P_i$ as $L\to\infty$.
\end{enumerate}
\end{lemma}

\begin{proof}
(a) $C_i$ contains $I_2$, so $(A_i,C_i)$ is observable. The determinant of $(b_i,\ A_ib_i)$ is $-2\pi T_s$ for $i=r$ and $-2\pi T_s(1-\rho)^2$ for $i=g$, which are nonzero, so $(A_i,b_i)$ is controllable. Under these two conditions, the stabilizing solution exists and is unique, and $\Ric_i^L(\Pi)\to P_i$ for every $\Pi\in\Sym^2_+$~\cite[Sec.~4.4]{anderson1979}. To show $P_i\succ0$, drop the index $i$ and let $P^f=P-PC^\top(CPC^\top+R)^{-1}CP$, so that $P=AP^fA^\top+Q$. By the Woodbury identity, $P^f=P^{1/2}(I+P^{1/2}C^\top R^{-1}CP^{1/2})^{-1}P^{1/2}$, so $P^f$ and $P$ have the same null space $\mathcal V$. If $v\in\mathcal V$, then $0=v^\top Pv=(A^\top v)^\top P^f(A^\top v)+q(b^\top v)^2$, where $q=q_w$ or $q_d$; hence $b^\top v=0$ and $A^\top v\in\mathcal V$. So $\mathcal V$ is invariant under $A^\top$ and orthogonal to $b$. If $\mathcal V\neq\{0\}$, it contains a (possibly complex) eigenvector $v$ of $A^\top$ with $b^\top v=0$, which contradicts controllability by the Popov--Belevitch--Hautus test. Hence $\mathcal V=\{0\}$.
(b) Before the window, the record consists of linear measurements of the states in independent Gaussian noise, and the commands are functions of the record available when they are applied. For such systems, the conditional law of the state given the record is Gaussian, and its covariance obeys the Riccati recursion started at $\Pi_i$, which involves neither the data nor the commands~\cite[Ch.~3]{anderson1979},~\cite[Ch.~7]{kumar1986}. The matrices $A$, $Q$, $C$, $R$ of the two sides together are block diagonal, so the conditional covariance is block diagonal; for jointly Gaussian errors, this is conditional independence. The limit follows from (a).
\end{proof}

\begin{assumption}[Steady state at the audit]\label{as:ss}
For each $i\in\{r,g\}$, $x_{i,-L}\sim\N(\bar x_i,P_i)$ for some $\bar x_i\in\R^2$, where $P_i$ is the matrix of Lemma~\ref{lem:kalman}(a).
\end{assumption}

Since $P_i=\Ric_i(P_i)$, Assumption~\ref{as:ss} and Lemma~\ref{lem:kalman}(b) imply that the defender's prediction error of $x_{i,0}$ has covariance exactly $P_i$, for every $L\in\Z_{\ge0}$. Assumption~\ref{as:ss} thus states that the defender has watched the system long enough for its predictions to be in steady state when the window opens; by Lemma~\ref{lem:kalman}(b), it is the limit of a long observation from any initial covariance. We write
\[
s_i=1/\sqrt{(P_i^{-1})_{22}}\in(0,\infty)\quad\text{(Hz)}.
\]
For a Gaussian vector with covariance $P_i$, $s_i^2$ is the variance of the second component given the first: $s_i$ is the defender's uncertainty about the frequency of side $i$ at frame $0$, given its phase.

Under~\ref{as:lin}--\ref{as:ss}, the model defines the law $\Pb^\tau_\lambda$ of the record $\mathcal Y$ for every $\tau\in\R^{2N}$, not only for $\tau\in\A_N$, and we use the evidence $\Ev_\lambda(\tau)$ of Definition~\ref{def:game} for every such $\tau$. For one clock, we write $\tau_i=(\tau_{i,1},\dots,\tau_{i,N})\in\R^N$, so that $\tau=(\tau_r,\tau_g)$.

\section{The Evidence of an Attack}\label{sec:evidence}

This section computes the payoff. The evidence vanishes if the attack changes nothing that the defender sees, and it is large if the attacked record looks unusual. We make this precise, starting from a single measurement.

\subsection{Warm-up: one measurement}

Suppose the defender sees a single number that is $\N(0,\sigma^2)$ without attack and $\N(\mu,\sigma^2)$ with it, where $\mu\in\R$ and $\sigma\in(0,\infty)$. By (F3), the evidence is $\mu^2/(2\sigma^2)$: half the squared shift, counted in noise standard deviations. For $\alpha\in(0,\frac12)$ and $\mu>0$, the best test with false-alarm probability $\alpha$ raises an alarm when the number exceeds $z_{1-\alpha}\sigma$, and it detects the attack with probability $\Phi(\mu/\sigma-z_{1-\alpha})$. For example, a shift of $3\sigma$ has evidence $4.5$\,nats; at $\alpha=10^{-3}$, where $z_{1-\alpha}=3.09$, it is detected with probability $0.46$. The full problem is this example in vector form, as we now show.

\subsection{The evidence is a quadratic form}

\emph{The window record in vector form.} Collect in $\eta$ everything that the defender cannot predict about normal operation, the \emph{nuisance}:
\[
\begin{aligned}
\eta=\big(&x_{r,0}-\hat x_{r,0},\,x_{g,0}-\hat x_{g,0},\\
&w_0,\dots,w_{N-2},\,d_0,\dots,d_{N-2},\,\delta\big),
\end{aligned}
\]
where $\delta$ lists the offsets $\delta_i$ of the installed oscillators, so that $\eta\in\R^p$ with $p=2N+2+J_o$. Collect in $\varepsilon\in\R^m$ the measurement noises $v_{i,k}$, $v'_{i,k}$, $n_{i,k}$ and $e_k$ of the frames $k\in\Kw_N$, and the commands of the window in $u=(u_0,\dots,u_{N-2})\in\R^{N-1}$. Iterating~\eqref{eq:util}--\eqref{eq:ref} from frame $0$ gives
\begin{equation}\label{eq:stack}
Y=\hat Y+F\eta+Hu+G\tau+\varepsilon ,
\end{equation}
where:
\begin{itemize}
\item $\hat Y\in\R^m$ is the prediction of the record, its response to $\hat x_{r,0}$ and $\hat x_{g,0}$. It is a function of $\Yp$ alone, the same with and without attack.
\item $F\in\R^{m\times p}$ and $H\in\R^{m\times(N-1)}$ give the responses of the record to the nuisance and to the commands. Since $u_k$ affects the island only from frame $k+1$ on, the rows of $H$ for frame $k$ involve only $u_0,\dots,u_{k-1}$.
\item $G\in\R^{m\times2N}$ gives the \emph{signature}: in frame $k\in\Kw_N$, the rows of $\pth_{i,k}$, $\pnu_{i,k}$, $o_{i,k}$ and $\tref_k$ carry $-\omega_0\tau_{i,k}$, $-(f_0/T_s)(\tau_{i,k+1}-\tau_{i,k})$, $\tau_{i,k}$ and $0$, with $\tau_{i,0}=0$.
\item Given $\Yp$, $\eta\sim\N(0,P)$ and $\varepsilon\sim\N(0,R)$ are independent, with
\[
P=\mathrm{diag}\big(P_r,P_g,q_wI_{N-1},q_dI_{N-1},\Sigma_\delta\big)\in\Sym^p_{++},
\]
where $\Sigma_\delta$ is the diagonal matrix of the $\sigma_i^2$ of the installed oscillators, and $R\in\Sym^m_{++}$ is diagonal with entries $\sigma_\theta^2$, $\sigma_\nu^2$, $\sigma_o^2$ and $\sigma_e^2$. This follows from Lemma~\ref{lem:kalman}(b) and Assumptions~\ref{as:gauss}--\ref{as:ss}, since $w_k$, $d_k$, $\delta_i$ and the noises of the window are independent of $\Yp$.
\end{itemize}

\begin{theorem}[Evidence as explanation cost]\label{thm:evidence}
Let~\ref{as:lin}--\ref{as:ss} hold. For every $L\in\Z_{\ge0}$, every admissible controller, every design and every $\tau\in\R^{2N}$:
\begin{enumerate}
\item[(a)] \emph{Controller invariance.} The evidence is the quadratic form
\begin{equation}\label{eq:quad}
\Ev_\lambda(\tau)=\tfrac12\tau^\top M\tau,\qquad M=G^\top\Psi^{-1}G,
\end{equation}
with $\Psi=R+FPF^\top$, where $\Psi\in\Sym^m_{++}$ and $M\in\Sym^{2N}_{++}$. In particular, $\Ev_\lambda(\tau)$ is finite, it depends neither on $L$ nor on the controller, and it is positive for $\tau\neq0$.
\item[(b)] \emph{Explanation cost} (standard information form~\cite{anderson1979,scharf1994}).
\begin{equation}\label{eq:explain}
2\,\Ev_\lambda(\tau)=\min_{\eta\in\R^p}\Big\{\|G\tau-F\eta\|^2_{R^{-1}}+\|\eta\|^2_{P^{-1}}\Big\},
\end{equation}
and the minimum is attained at the unique point $\hat\eta=(P^{-1}+F^\top R^{-1}F)^{-1}F^\top R^{-1}G\tau$.
\item[(c)] \emph{Bounding duality.} Explanations bound from above: for every $\eta\in\R^p$, $\Ev_\lambda(\tau)\le\frac12\big(\|G\tau-F\eta\|^2_{R^{-1}}+\|\eta\|^2_{P^{-1}}\big)$, the \emph{cost} of $\eta$. Statistics bound from below: for every $c\in\R^m\setminus\{0\}$, $\Ev_\lambda(\tau)\ge(c^\top G\tau)^2/(2c^\top\Psi c)$.
\item[(d)] \emph{Additivity.} With the parts of the record defined in Section~\ref{sec:additivity},
\[
\Ev_\lambda(\tau)=\underbrace{\Ev_{y_r}(\tau_r)+\Ev_{o_r}(\tau_r)}_{\Ev_r(\tau_r)}+\underbrace{\Ev_{y_g}(\tau_g)+\Ev_{o_g}(\tau_g)}_{\Ev_g(\tau_g)},
\]
each term is given by (a)--(b) with $F,G,P,R$ restricted to that part, $M=\mathrm{diag}(M_r,M_g)$, and $\Ev_i(\tau_i)=\frac12\tau_i^\top M_i\tau_i$ with $M_i\in\Sym^N_{++}$.
\end{enumerate}
\end{theorem}

\inwords This is the warm-up formula, with the shift $\mu$ replaced by the signature $G\tau$ and the variance $\sigma^2$ by $\Psi$, the covariance of everything that the defender cannot predict: measurement noise ($R$) and nuisance ($FPF^\top$). The controller does not appear. Its commands are computed from the data, so the defender can subtract their effect; the controller can neither hide the attack nor reveal it. This invariance concerns the evidence only. The damage, the phase difference that the attack causes, depends on the controller through the tracking error of Section~\ref{sec:damage}.

\begin{proof}[Proof of Theorem~\ref{thm:evidence}(a)]
By (F1), $\Ev_\lambda(\tau)$ is the divergence of the laws of $\Yp$ plus the expected divergence of the conditional laws of $Y$ given $\Yp$. The first term is zero: before frame $0$ the signature vanishes, since $\tau_{i,k}=0$ for $k\le0$, so $\Yp$ has the same law with and without attack. Fix a value of $\Yp$. By admissibility, $u_k$ is a measurable function of $\Yp$ and $Y_0,\dots,Y_k$. Consider the map $Y\mapsto Z=Y-Hu$. Its inverse is computed frame by frame: $Y_0=Z_0$, because the window commands first affect frame~$1$, and $Y_k=Z_k+(Hu)_k$, where $(Hu)_k$ depends on $u_0,\dots,u_{k-1}$ and hence on $Y_0,\dots,Y_{k-1}$. Both the map and its inverse are Borel measurable, so by (F2) the conditional divergence of $Y$ equals that of $Z$. By~\eqref{eq:stack}, $Z=\hat Y+F\eta+G\tau+\varepsilon$, whose conditional law is $\N(\hat Y+G\tau,\Psi)$ with attack and $\N(\hat Y,\Psi)$ without, where $\Psi\succeq R\succ0$. By (F3), the conditional divergence is $\frac12\tau^\top G^\top\Psi^{-1}G\tau$ for every value of $\Yp$, which gives~\eqref{eq:quad}. Finally, $G$ is injective: if $G\tau=0$, the phase rows of frames $1,\dots,N-1$ give $\tau_{i,1}=\dots=\tau_{i,N-1}=0$, and then the frequency row of frame $N-1$ gives $\tau_{i,N}=0$. With $\Psi^{-1}\succ0$, $\tau^\top M\tau=\|G\tau\|^2_{\Psi^{-1}}>0$ for $\tau\neq0$.
\end{proof}

\subsection{The evidence is the cost of the cheapest innocent explanation}

\inwords Identity~\eqref{eq:explain} is the standard information form of a Gaussian divergence with a Gaussian nuisance, the objective of Kalman smoothing and, without the prior term, of the matched subspace detector~\cite{anderson1979,scharf1994}; what is new is what it yields for timing attacks. Theorem~\ref{thm:evidence}(b) is the lens for the rest of the paper. An \emph{explanation} $\eta$ is an account of normal operation: a value of the nuisance, such as a grid-frequency path, a load path or an oscillator offset. It reproduces the signature $G\tau$ up to a misfit $G\tau-F\eta$, which must then be blamed on measurement noise. Half the bracket in~\eqref{eq:explain} is the \emph{cost} of the explanation: how unlikely the nuisance is, plus how unlikely the misfit is, each measured as in the warm-up by half its squared size in standard deviations. The evidence is the cost of the cheapest explanation, $\hat\eta$. An attack is stealthy precisely when normal operation can account for it cheaply.

\begin{proof}[Proof of Theorem~\ref{thm:evidence}(b)]
Since $P\succ0$, the bracket is a strictly convex quadratic function of $\eta$, and setting its gradient to zero gives $\hat\eta$. Substituting $\hat\eta$, the minimum equals $(G\tau)^\top\big(R^{-1}-R^{-1}F(P^{-1}+F^\top R^{-1}F)^{-1}F^\top R^{-1}\big)G\tau$, and by the Woodbury identity the matrix in parentheses is $(R+FPF^\top)^{-1}=\Psi^{-1}$. Part (a) completes the proof.
\end{proof}

\begin{proof}[Proof of Theorem~\ref{thm:evidence}(c)]
The upper bound follows from~\eqref{eq:explain}. For the lower bound, by the Cauchy--Schwarz inequality,
\[
\begin{aligned}
(c^\top G\tau)^2&=\big((\Psi^{1/2}c)^\top(\Psi^{-1/2}G\tau)\big)^2\\
&\le\|\Psi^{1/2}c\|^2\,\|\Psi^{-1/2}G\tau\|^2=(c^\top\Psi c)\,\tau^\top M\tau,
\end{aligned}
\]
and $c^\top\Psi c>0$ because $\Psi\succ0$.
\end{proof}

In the lower bound of (c), $c^\top Z$ is a linear statistic of the record from which the known effect of the commands has been removed. Without attack it has variance $c^\top\Psi c=\mathrm{Var}\big(c^\top(F\eta+\varepsilon)\big)$, and the attack shifts its mean by $c^\top G\tau$. The lower bound thus says that the whole record reveals at least the warm-up evidence of any one statistic. Every explanation gives an upper bound, every statistic a lower bound, and the two meet at the cheapest explanation. This duality is what makes the safe audit periods of Section~\ref{sec:results} certifiable.

\emph{Example.} Let the utility clock drift at the constant rate $\gamma\in\R$ that brings it to $\tc$ after 60\,s, namely $\gamma=9.3\times10^{-6}$. By~\eqref{eq:pmu}, the utility PMU then reports a frequency that is $f_0\gamma=0.46$\,mHz too low and a phase that falls accordingly. A grid frequency $0.46$\,mHz lower than predicted at frame $0$ explains this perfectly. With the uncertainty $s_r=0.24$\,mHz of Section~\ref{sec:results}, this explanation costs $\frac12(0.46/0.24)^2\approx1.8$\,nats, so the evidence is at most $1.8$\,nats. By Proposition~\ref{prop:detect} below, this is insufficient for $50\%$ detection at $\alpha=10^{-3}$: even the test matched to this attack detects it with probability at most about $12\%$.

\subsection{Additivity}\label{sec:additivity}

Write $Z=Y-Hu$ as in the proof of Theorem~\ref{thm:evidence}(a), and split it into four \emph{parts}: $Z_{y_r}$, the utility PMU reports together with the reference, if installed; $Z_{y_g}$, the island PMU reports; and $Z_{o_r}$ and $Z_{o_g}$, the holdover readings, if installed. Only $Z_{y_g}$ differs from the raw data, since the commands affect only the island. The \emph{evidence of a part}, $\Ev_{y_r}$, $\Ev_{y_g}$, $\Ev_{o_r}$ or $\Ev_{o_g}$, is the divergence between the conditional laws of that part given $\Yp$ with and without attack; it is zero for an absent monitor.

\begin{proof}[Proof of Theorem~\ref{thm:evidence}(d)]
The four parts have independent nuisances, $(x_{r,0}-\hat x_{r,0},w)$, $(x_{g,0}-\hat x_{g,0},d)$, $\delta_r$ and $\delta_g$ (Lemma~\ref{lem:kalman}(b)), and independent noises. After reordering, $F$, $P$ and $R$ are therefore block diagonal over the parts, and so is $\Psi$. The signature of $\tau_r$ enters only the first and third parts, and that of $\tau_g$ only the second and fourth. Given $\Yp$, the four parts of $Z$ are thus independent Gaussian vectors, and the divergence of independent vectors is the sum of their divergences, by (F1). Each $M_i$ is positive definite by the injectivity argument of Theorem~\ref{thm:evidence}(a) applied to the PMU of side $i$.
\end{proof}

\subsection{What each monitor can explain}\label{sec:regimes}

We now apply the bounding principle to one clock, with $\tau_{i,0}=0$ and $\gamma_{i,k}=(\tau_{i,k+1}-\tau_{i,k})/T_s$. Two quantities appear: $S_N=T_s^2\sum_{k\in\Kw_N}k^2=T_s^2(N-1)N(2N-1)/6$ (s$^2$), and the variance $\sigma_{\theta e}^2=\sigma_\theta^2+\sigma_e^2$ of $\pth_{r,k}-\tref_k$ without attack. For the island and $N\ge2$, let
\[
\begin{aligned}
V^g_N={}&(2\pi)^2q_d\sum_{k=0}^{N-2}\big(T_s-\beta\rho^k\big)^2\\
&+\big(2\pi T_g(1-\rho^{N-1})\big)^2(P_g)_{22}+2\sigma_\theta^2
\end{aligned}
\]
(rad$^2$), and $N_g=\big((T_g/T_s)^2(P_g)_{22}+\sigma_\theta^2/(2\pi^2T_s^2)\big)/q_d\in(0,\infty)$, a number of frames.

\begin{lemma}[Bounds for every profile]\label{lem:bounds}
Let~\ref{as:lin}--\ref{as:ss} hold, $N\in\Nat$ with $N\ge2$, and $\gamma\in\R$.
\begin{enumerate}
\item[(a)] \emph{Utility PMU, no reference:} the ramp $\tau_{r,k}=\gamma kT_s$ has $\Ev_{y_r}(\tau_r)\le(f_0\gamma)^2/(2s_r^2)$.
\item[(b)] \emph{Island PMU:} every $\tau_g\in\R^N$ has
\[
\Ev_{y_g}(\tau_g)\ge\frac{\omega_0^2\tau_{g,N-1}^2}{2V^g_N}\ge\frac{f_0^2\tau_{g,N-1}^2}{2T_s^2q_d(N-1+N_g)},
\]
and the ramp $\tau_{g,k}=\gamma kT_s$ has $\Ev_{y_g}(\tau_g)\le\frac12(f_0\gamma)^2\big((N-1)/q_d+1/s_g^2\big)$.
\item[(c)] \emph{Holdover check $i$:} the ramp $\tau_{i,k}=\gamma kT_s$ has $\Ev_{o_i}(\tau_i)=\gamma^2/\big(2(\sigma_i^2+\sigma_o^2/S_N)\big)$, and every $\tau_i\in\R^N$ has $\Ev_{o_i}(\tau_i)\ge\tau_{i,N-1}^2/\big(2(\sigma_i^2(N-1)^2T_s^2+\sigma_o^2)\big)$.
\item[(d)] \emph{Utility PMU with reference:} every $\tau_r\in\R^N$ has $\Ev_{y_r}(\tau_r)\ge\frac{\omega_0^2}{2\sigma_{\theta e}^2}\sum_{k\in\Kw_N}\tau_{r,k}^2$ and
\[
\Ev_{y_r}(\tau_r)\le\frac{\omega_0^2}{2\sigma_\theta^2}\sum_{k\in\Kw_N}\tau_{r,k}^2+\frac{f_0^2}{2\sigma_\nu^2}\sum_{k\in\Kw_N}\gamma_{r,k}^2.
\]
\end{enumerate}
\end{lemma}

\begin{proof}
We apply Theorem~\ref{thm:evidence}(c) to each part (Theorem~\ref{thm:evidence}(d)).

(a) Take the explanation that shifts $x_{r,0}$ by $(0,-f_0\gamma)^\top$ and keeps $w=0$. By~\eqref{eq:util}, the grid frequency is then lower by $f_0\gamma$ in every frame and the grid phase lower by $2\pi T_sf_0\gamma k=\omega_0\tau_{r,k}$, which is exactly the signature~\eqref{eq:pmu}. The misfit is zero and the cost is $\frac12\|(0,f_0\gamma)^\top\|^2_{P_r^{-1}}=\frac12(f_0\gamma)^2(P_r^{-1})_{22}=(f_0\gamma)^2/(2s_r^2)$.

(b) Lower bound: take the statistic $c^\top Z=\tilde Z_{N-1}-\tilde Z_0$, where $\tilde Z_k$ is the phase entry of $Z_{y_g}$ in frame $k$. Its mean shift is $-\omega_0(\tau_{g,N-1}-\tau_{g,0})=-\omega_0\tau_{g,N-1}$. Without attack and with the commands removed, iterating~\eqref{eq:isl} from frame $0$ gives
\[
\begin{aligned}
\theta_{g,N-1}-\theta_{g,0}={}&2\pi T_g(1-\rho^{N-1})\,\nu_{g,0}\\
&+2\pi\sum_{l=0}^{N-2}\big(T_s-\beta\rho^{N-2-l}\big)d_l ,
\end{aligned}
\]
in which only the prediction error of $\nu_{g,0}$, of variance $(P_g)_{22}$, and the independent loads are random. Adding the noises $v_{g,N-1}-v_{g,0}$ gives $c^\top\Psi c=V^g_N$, which proves the first inequality. For the second, $0<T_s-\beta\rho^j<T_s$ and $(1-\rho^{N-1})^2<1$ give $V^g_N\le(2\pi T_s)^2q_d(N-1+N_g)$, and $\omega_0^2/(2\pi)^2=f_0^2$.
Upper bound: take the explanation that shifts $x_{g,0}$ by $(0,-f_0\gamma)^\top$ and sets $d_k=-f_0\gamma$ for $k\le N-2$. By~\eqref{eq:isl}, the island frequency then stays lower by $f_0\gamma$, and its phase falls by $2\pi\big(\beta+(T_s-\beta)\big)f_0\gamma=\omega_0\gamma T_s$ per frame, which reproduces the signature. The cost is $\frac12(f_0\gamma)^2\big(1/s_g^2+(N-1)/q_d\big)$.

(c) This part is a linear regression: given $\Yp$, $Z_{o_i}=(\tau_{i,0},\dots,\tau_{i,N-1})^\top+\delta_it+n_i$ with $t=(0,T_s,\dots,(N-1)T_s)^\top$, so $\Psi_{o_i}=\sigma_o^2I_N+\sigma_i^2tt^\top$. For the ramp, the signature is $\gamma t$, and by the Sherman--Morrison formula, $\frac12\gamma^2t^\top\Psi_{o_i}^{-1}t=\gamma^2\|t\|^2/\big(2(\sigma_o^2+\sigma_i^2\|t\|^2)\big)$ with $\|t\|^2=S_N$. For the lower bound, take the statistic $o_{i,N-1}$, whose mean shift is $\tau_{i,N-1}$ and whose variance is $\sigma_i^2(N-1)^2T_s^2+\sigma_o^2$.

(d) Lower bound: if $\tau_{r,k}=0$ for all $k\in\Kw_N$ there is nothing to prove; otherwise take the statistic $\sum_{k\in\Kw_N}\tau_{r,k}(\pth_{r,k}-\tref_k)$. Since $\pth_{r,k}-\tref_k=-\omega_0\tau_{r,k}+v_{r,k}-e_k$, the grid phase cancels; the mean shift is $-\omega_0\sum_k\tau_{r,k}^2$ and the variance $\sigma_{\theta e}^2\sum_k\tau_{r,k}^2$. Upper bound: take the explanation $\eta=0$, for which the misfit is the whole signature, of cost $\frac12\sum_k\big(\omega_0^2\tau_{r,k}^2/\sigma_\theta^2+f_0^2\gamma_{r,k}^2/\sigma_\nu^2\big)$; the reference rows carry no signature.
\end{proof}

We now specialize to the simplest attack on one clock, a \emph{ramp} that reaches the bias $a\in\R$ at the audit with constant drift: $\tau_k=ak/N$ for $k\in\{0,\dots,N\}$, so $\gamma=a/T$ with $T=NT_s$.

\begin{proposition}[Three regimes]\label{prop:regimes}
Let~\ref{as:lin}--\ref{as:ss} hold, $N\in\Nat$ with $N\ge2$, $T=NT_s$ and $a\in\R$. The ramp that reaches the bias $a$ in time $T$ reveals the following evidence:
\begin{enumerate}
\item[(a)] to the utility PMU without reference, at most
\[
\frac{f_0^2a^2}{2s_r^2T^2};
\]
\item[(b)] to the island PMU, between $\ell_NK_g$ and $u_NK_g$, where
\[
K_g=\frac{f_0^2a^2}{2q_dT_sT},\qquad \ell_N=\frac{(N-1)^2}{N(N-1+N_g)},
\]
and $u_N=1-(1-q_d/s_g^2)/N$; both $\ell_N$ and $u_N$ tend to $1$ as $N\to\infty$;
\item[(c)] to holdover check $i$, exactly
\[
\frac{a^2}{2T^2(\sigma_i^2+\sigma_o^2/S_N)};
\]
\item[(d)] to the utility PMU with reference, at least
\[
\frac{\omega_0^2a^2(N-1)(2N-1)}{12N\sigma_{\theta e}^2}\approx\frac{\omega_0^2a^2T}{6T_s\sigma_{\theta e}^2}\quad\text{for large }N.
\]
\end{enumerate}
\end{proposition}

\begin{proof}
Set $\gamma=a/T$ in Lemma~\ref{lem:bounds}. (a) and (c) are immediate. (b) With $\tau_{g,N-1}=a(N-1)/N$, the lower bound of Lemma~\ref{lem:bounds}(b) is $f_0^2a^2(N-1)^2/\big(2N^2T_s^2q_d(N-1+N_g)\big)$, which is $\ell_NK_g$; the upper bound is $f_0^2a^2\big((N-1)/q_d+1/s_g^2\big)/(2N^2T_s^2)$, which is $u_NK_g$. (d) $\sum_{k\in\Kw_N}\tau_k^2=\gamma^2S_N$, and $\gamma^2S_N=a^2(N-1)(2N-1)/(6N)$.
\end{proof}

\begin{table}[t]
\caption{What each monitor can blame a drift on (Proposition~\ref{prop:regimes}), and how the evidence of a ramp that reaches the bias $a$ in time $T$ scales; $O(\cdot)$ marks an upper bound}\label{tab:regimes}
\centering\footnotesize
\setlength{\tabcolsep}{3pt}
\begin{tabular}{@{}lllc@{}}
\toprule
Monitor & Innocent explanation & Paid & Evidence\\
\midrule
utility PMU & lower grid frequency & once & $O(a^2/T^2)$\\
holdover check & slower oscillator & once & $\propto a^2/T^2$\\
island PMU & heavier load & every frame & $\propto a^2/T$\\
trusted reference & none & -- & $\propto a^2\,T$\\
\bottomrule
\end{tabular}
\end{table}

\inwords Table~\ref{tab:regimes} summarizes the result, and Fig.~\ref{fig:results}(a) shows it numerically. The utility PMU blames the drift on a grid frequency lower than predicted, as in the example above, and the holdover check blames it on an oscillator that runs slow. Both explanations are paid for once, for the whole window, and the slower the attack, the smaller the offset they need. For the utility PMU, Proposition~\ref{prop:regimes}(a) is only an upper bound; Fig.~\ref{fig:results}(a) shows the same decay numerically in the case study. The island PMU blames the drift on a heavier load, which the explanation needs in every frame, because the load is white noise. The trusted reference measures the grid phase without GNSS and cannot blame the drift on anything. Slowing the attack down therefore helps the attacker against the first three monitors and hurts it against the fourth. Parts (a) and (c) even have the same form: the utility PMU reveals no more than an oscillator of spread $s_r/f_0$. (The holdover evidence scales as $a^2/T^2$ once $S_N\gg\sigma_o^2/\sigma_i^2$. Since $S_N\approx T^3/(3T_s)$, the two are equal at $T\approx(3T_s\sigma_o^2/\sigma_i^2)^{1/3}$: $0.06$\,s for the TCXO of Section~\ref{sec:results}, $1.3$\,s for the OCXO and $28$\,s for the rubidium standard. The scaling holds well beyond these times.)

\subsection{What evidence buys}

For $\tau\in\R^{2N}$, let $\Lambda=\log\frac{d\Pb^\tau_\lambda}{d\Pb^0_\lambda}(\mathcal Y)$ be the log-likelihood ratio of the record. A \emph{test} with false-alarm probability $\alpha$ is a measurable map $\phi$ from records to $[0,1]$ (the probability of an alarm) with $\Ex^0[\phi]\le\alpha$, and its probability of detection is $\Ex^\tau[\phi]$, where $\Ex^0$ and $\Ex^\tau$ are expectations under $\Pb^0_\lambda$ and $\Pb^\tau_\lambda$. A detector that knows the profile $\tau$ but not when the attack starts is described by the \emph{Shiryaev--Roberts} (SR) procedure~\cite{tartakovsky2014} matched to $\tau$; it scans the onset, not the shape. For $n\in\Kw_N$ and an onset $s\in\{0,\dots,n\}$, the \emph{delayed profile} has biases $\tau_{i,k-s}$ at frames $k\ge s$ and $0$ before. The oscillators stay synchronized at frame $0$, and the record up to frame $n$ involves only the entries $\tau_{i,k}$ with $k\le n-s+1\le N$, so $\tau$ needs no continuation beyond the window. Let $\Lambda_{s,n}$ be the likelihood ratio, given $\Yp$, of the record up to frame $n$ under the delayed profile against no attack, let $R_n=\sum_{s=0}^n\Lambda_{s,n}$, and raise the alarm at $T_B=\inf\{n\in\Kw_N:R_n\ge B\}$, for a threshold $B\in(1,\infty)$. After an alarm, the procedure restarts: its sum then runs only over the onsets after the alarm.

\begin{proposition}[Detection]\label{prop:detect}
Let~\ref{as:lin}--\ref{as:ss} hold, $\tau\in\R^{2N}\setminus\{0\}$ and $\Ev=\Ev_\lambda(\tau)$.
\begin{enumerate}
\item[(a)] For every $\alpha\in(0,\frac12)$, the most powerful test with false-alarm probability $\alpha$ detects $\tau$ with probability $\Phi\big(\sqrt{2\Ev}-z_{1-\alpha}\big)$. This is at least $\frac12$ if and only if $\Ev\ge h_Z:=\frac12z_{1-\alpha}^2$; if $\Ev<h_Z$, every test with false-alarm probability $\alpha$ misses $\tau$ with probability above $\frac12$.
\item[(b)] For every $B\in(1,\infty)$ and $n\in\Kw_N$, the SR detector matched to $\tau$ raises in frames $0,\dots,n$ an expected number of false alarms of at most $(n+1)/B$, that is, on average at most one per $B$ frames; in particular, $\Pb^0_\lambda(T_B\le n)\le(n+1)/B$. It detects $\tau$ by frame $N-1$ with probability at least $\Phi\big((\Ev-\log B)/\sqrt{2\Ev}\big)$, which is at least $\frac12$ if $\Ev\ge h_B:=\log B$.
\end{enumerate}
\end{proposition}

Part (a) is the warm-up with $\mu/\sigma=\sqrt{2\Ev}$; it assumes that the test knows the profile, including when the attack starts. Part (b) relaxes only the onset. The SR detector still knows the shape of the profile, and it must set a higher threshold to keep false alarms rare during continuous operation; $h_B$ prices the unknown onset, not an unknown shape. For $\alpha=10^{-3}$, $h_Z=4.77$\,nats; one false alarm per day at 50 frames per second gives $B=4.3\times10^6$ and $h_B=15.3$\,nats. Both thresholds aim at a detection probability of $\frac12$ for the matched detector. A target $p\in[\frac12,1)$ for the test of part (a) needs $h_p=\frac12(z_{1-\alpha}+z_p)^2$, for example $h_p=9.56$\,nats for $p=0.9$ at $\alpha=10^{-3}$.

\begin{proof}
By the proof of Theorem~\ref{thm:evidence}(a), the likelihood ratio is that of $Z$ given $\Yp$, so $\Lambda=(G\tau)^\top\Psi^{-1}(Z-\hat Y)-\Ev$. Since $Z-\hat Y\sim\N(0,\Psi)$ without attack and $\N(G\tau,\Psi)$ with it, whatever $\Yp$, $\Lambda\sim\N(-\Ev,2\Ev)$ under $\Pb^0_\lambda$ and $\Lambda\sim\N(\Ev,2\Ev)$ under $\Pb^\tau_\lambda$.
(a) By the Neyman--Pearson lemma, the most powerful test raises an alarm when $\Lambda>-\Ev+\sqrt{2\Ev}\,z_{1-\alpha}$, which has false-alarm probability $\alpha$. Its probability of detection is $\Pb\big(\N(\Ev,2\Ev)>-\Ev+\sqrt{2\Ev}z_{1-\alpha}\big)=\Phi(\sqrt{2\Ev}-z_{1-\alpha})$, which is at least $\frac12$ if and only if $\sqrt{2\Ev}\ge z_{1-\alpha}$; here $z_{1-\alpha}>0$ because $\alpha<\frac12$.
(b) An attack that starts at frame $s$ does not change the law of the record before frame $s$. Hence, under $\Pb^0_\lambda$ and with respect to the record up to frame $n$, $(\Lambda_{s,n})_{n\ge s}$ is a martingale, and $\Ex^0[\Lambda_{n+1,n+1}\mid\text{record up to }n]=1$. Let $\tilde R_n$ be the statistic of the current cycle, so that $\tilde R_n=R_n$ up to the first alarm, and let $X_n=\tilde R_n+\sum_a\tilde R_a-(n+1)$, the sum running over the alarm frames $a<n$. If $n$ is not an alarm frame, $X_{n+1}-X_n=\sum_s(\Lambda_{s,n+1}-\Lambda_{s,n})+\Lambda_{n+1,n+1}-1$, the sum running over the onsets of the current cycle; if it is, $X_{n+1}-X_n=\Lambda_{n+1,n+1}-1$. Both increments have conditional mean zero, and $\Ex^0[X_0]=\Ex^0[\Lambda_{0,0}]-1=0$, so $\Ex^0[X_n]=0$. Since $\tilde R_a\ge B$ at every alarm frame $a$, $\tilde R_n\ge0$, and $\tilde R_n\ge B$ if $n$ is itself an alarm frame, $B$ times the expected number of alarms in frames $0,\dots,n$ is at most $\Ex^0[X_n]+n+1=n+1$. The first alarm is at $T_B$, so $\Pb^0_\lambda(T_B\le n)\le(n+1)/B$. For detection, $R_{N-1}\ge\Lambda_{0,N-1}=e^\Lambda$, so $\Pb^\tau_\lambda(T_B\le N-1)\ge\Pb^\tau_\lambda(\Lambda\ge\log B)=\Phi\big((\Ev-\log B)/\sqrt{2\Ev}\big)$.
\end{proof}

\section{The Attacker's Best Response}\label{sec:attacker}

Fix a design $\lambda$; we drop the index $\lambda$ throughout this section. The attacker solves $\min_{\tau\in\A_N}\Ev(\tau)$.

\begin{theorem}[Best response]\label{thm:br}
Let~\ref{as:lin}--\ref{as:ss} hold, $N\in\Nat$ with $2N\Delta\ge\tc$, and let $\ell\in\R^{2N}$ pick out the differential bias: $\ell^\top\tau=\tau_{r,N}-\tau_{g,N}$.
\begin{enumerate}
\item The stealthiest attack exists and is the unique solution $\tau^\star$ of the strictly convex quadratic program (QP) $\min\{\frac12\tau^\top M\tau:\tau\in\A_N\}$; thus $\Es(N)=\Ev(\tau^\star)$.
\item $\Es(N)\ge(\tc)^2/(2\ell^\top M^{-1}\ell)$, with equality if $\hat\tau=\tc M^{-1}\ell/(\ell^\top M^{-1}\ell)$ satisfies the rate limit (R), in which case $\tau^\star=\hat\tau$.
\item With $M_i$ as in Theorem~\ref{thm:evidence}(d), let $\kappa_i=1/\big(2(M_i^{-1})_{NN}\big)$, a number in $(0,\infty)$. Then $\kappa_ia^2=\min\{\Ev_i(\tau_i):\tau_i\in\R^N,\ \tau_{i,N}=a\}$ for every $a\in\R$. If $\hat\tau$ satisfies (R), then
\begin{equation}\label{eq:harmonic}
\Es(N)=\frac{(\tc)^2}{1/\kappa_r+1/\kappa_g},\qquad \frac{a_r^\star}{a_g^\star}=\frac{\kappa_g}{\kappa_r},
\end{equation}
where $a_r^\star=\tau^\star_{r,N}$ and $a_g^\star=-\tau^\star_{g,N}$ split $\tc$ between the clocks: $a_r^\star+a_g^\star=\tc$.
\end{enumerate}
\end{theorem}

\inwords $\kappa_i$ is the least evidence per unit squared bias of clock $i$ when the rate limit does not bind: biasing clock $i$ alone by $a$ costs $\kappa_ia^2$. Item~3 corrects the intuition that the attacker should drag only the weaker clock. If the island clock costs three times as much per unit squared bias as the utility clock, the attacker puts $75\%$ of the bias on the utility clock and $25\%$ on the island clock. It then pays only $0.75$ of the cost of dragging the utility clock alone, since $0.75^2+3\cdot0.25^2=0.75$. Like two resistors in parallel, the two clocks together are cheaper than either alone.

\begin{proof}
(1) By Lemma~\ref{lem:AN}, $\A_N$ is nonempty, convex and compact, and by Theorem~\ref{thm:evidence}(a), $\tau\mapsto\frac12\tau^\top M\tau$ is continuous and strictly convex. A minimizer therefore exists and is unique.
(2) The hyperplane $\mathcal H=\{\tau:\ell^\top\tau=\tc\}$ contains $\A_N$. On $\mathcal H$, the Lagrange condition $M\tau=\mu\ell$, $\mu\in\R$, and $\ell^\top\tau=\tc$ give the unique minimizer $\hat\tau$ and the value $(\tc)^2/(2\ell^\top M^{-1}\ell)$. The minimum over $\A_N\subseteq\mathcal H$ is at least this, with equality if $\hat\tau\in\A_N$, which holds if and only if $\hat\tau$ satisfies (R).
(3) The first claim is (2) applied to one clock without rate limit, with $\ell$ replaced by the last unit vector $\bar e=(0,\dots,0,1)^\top\in\R^N$: $\min\{\frac12\tau_i^\top M_i\tau_i:\bar e^\top\tau_i=a\}=a^2/(2\bar e^\top M_i^{-1}\bar e)=\kappa_ia^2$. By Theorem~\ref{thm:evidence}(d), $M=\mathrm{diag}(M_r,M_g)$ and $\ell=(\bar e,-\bar e)$, so $\ell^\top M^{-1}\ell=(M_r^{-1})_{NN}+(M_g^{-1})_{NN}=1/(2\kappa_r)+1/(2\kappa_g)$, which gives the value in~\eqref{eq:harmonic}. Moreover, $\hat\tau_{r,N}=\tc(M_r^{-1})_{NN}/(\ell^\top M^{-1}\ell)$ and $-\hat\tau_{g,N}=\tc(M_g^{-1})_{NN}/(\ell^\top M^{-1}\ell)$, whose ratio is $\kappa_g/\kappa_r$.
\end{proof}

The shape of the stealthiest attack on one clock depends on the monitor that watches it. For $N\in\Nat$ and $a\in\R$, let
\[
\A^1_N(a)=\big\{\tau_i\in\R^N:\ \tau_i\ \text{satisfies (R) and }\tau_{i,N}=a\big\},
\]
with $\tau_{i,0}=0$, be the one-clock profiles that obey the rate limit and reach the bias $a$; it is nonempty if and only if $|a|\le N\Delta$. For a part $X\in\{y_g,y_r\}$, let $\Ev^{\min}_X(N,a)=\min\{\Ev_X(\tau_i):\tau_i\in\A^1_N(a)\}$ be the least evidence that this part reveals; the minimum exists by compactness and continuity.

\begin{proposition}[Patience]\label{prop:shape}
Let~\ref{as:lin}--\ref{as:ss} hold, $a\in(\Delta,\infty)$ and $N\in\Nat$ with $N\Delta\ge a$.
\begin{enumerate}
\item[(a)] \emph{Island PMU:}
\[
\begin{aligned}
\frac{f_0^2(a-\Delta)^2}{2T_s^2q_d(N-1+N_g)}&\le\Ev^{\min}_{y_g}(N,a)\\
&\le\frac{f_0^2a^2}{2N^2T_s^2}\Big(\frac{N-1}{q_d}+\frac1{s_g^2}\Big).
\end{aligned}
\]
Both bounds decay as $1/N$, so patience pays: $\Ev^{\min}_{y_g}(N,a)\to0$ as $N\to\infty$. Their ratio, $u_N(N-1+N_g)/\big(N(1-\Delta/a)^2\big)$, tends to $(1-\Delta/a)^{-2}$, so the ramp is nearly optimal in value when $a\gg\Delta$. This does not show that the stealthiest profile is close to a ramp in shape.
\item[(b)] \emph{Utility PMU with reference:} let $\tau^\star_k=\big(a-(N-k)\Delta\big)_+$ for $k\in\{0,\dots,N\}$, which waits and then sprints at the rate limit for the last $a/\gb$ seconds. Let $n_a=\lfloor a/\Delta\rfloor$,
\[
S^\star=\sum_{k=1}^{n_a}(a-k\Delta)^2,\qquad \Gamma^\star=n_a\gb^2+\Big(\frac{a-n_a\Delta}{T_s}\Big)^2 .
\]
Then $\tau^\star\in\A^1_N(a)$ and
\[
\frac{\omega_0^2S^\star}{2\sigma_{\theta e}^2}\le\Ev^{\min}_{y_r}(N,a)\le\frac{\omega_0^2S^\star}{2\sigma_\theta^2}+\frac{f_0^2\Gamma^\star}{2\sigma_\nu^2},
\]
with $(a-\Delta)^3/(3\Delta)\le S^\star\le a^3/(3\Delta)$ and $n_a\gb^2\le\Gamma^\star\le(n_a+1)\gb^2$. Neither bound depends on $N$: the late sprint $\tau^\star$ achieves an $N$-independent upper bound, and waiting cannot drive the least evidence to zero.
\end{enumerate}
\end{proposition}

\inwords Against the island PMU, the attacker gains by spreading the drift as thinly as possible, because the load explanation is paid for in every frame. Against the trusted reference, every frame with a nonzero bias is charged, so keeping the bias at zero as long as possible and then sprinting comes within a constant factor of the least evidence, however long the window. Fig.~\ref{fig:results}(b) shows that, in the case study, the solutions of the QP of Theorem~\ref{thm:br} have these two shapes.

\begin{proof}
Every $\tau_i\in\A^1_N(a)$ satisfies $\tau_{i,k}\ge a-(N-k)\Delta$ for $k\in\{0,\dots,N\}$, by (R) and $\tau_{i,N}=a$.
(a) In particular $\tau_{i,N-1}\ge a-\Delta>0$, and Lemma~\ref{lem:bounds}(b) gives the lower bound for every admissible profile. The ramp $\tau_{i,k}=ak/N$ changes by $a/N\le\Delta$ per frame, so it is admissible, and Lemma~\ref{lem:bounds}(b) with $\gamma=a/(NT_s)$ gives the upper bound. Dividing the upper bound, $u_NK_g$ with $K_g$ as in Proposition~\ref{prop:regimes}, by the lower one gives the ratio.
(b) If $a-(N-k)\Delta>0$, then $\tau_{i,k}\ge\tau^\star_k>0$; otherwise $\tau^\star_k=0$. Either way $\tau_{i,k}^2\ge(\tau^\star_k)^2$, and Lemma~\ref{lem:bounds}(d) gives $\Ev_{y_r}(\tau_i)\ge\omega_0^2\sum_{k\in\Kw_N}(\tau^\star_k)^2/(2\sigma_{\theta e}^2)$. Substituting $N-k$ for $k$ and using $n_a\le a/\Delta\le N$, $\sum_{k\in\Kw_N}(\tau^\star_k)^2=\sum_{k=1}^N(a-k\Delta)_+^2=S^\star$. The profile $\tau^\star$ starts at $\tau^\star_0=(a-N\Delta)_+=0$, ends at $a$, and changes by $\Delta$ in each of the last $n_a$ frames, by $a-n_a\Delta\in[0,\Delta)$ in the frame before, and by $0$ otherwise. So $\tau^\star\in\A^1_N(a)$ and $\sum_k(\gamma^\star_k)^2=\Gamma^\star$, and Lemma~\ref{lem:bounds}(d) gives the upper bound. Finally, $x\mapsto(a-x\Delta)_+^2$ is nonincreasing, so comparing the sum with integrals gives $\int_1^{a/\Delta}(a-x\Delta)^2dx\le S^\star\le\int_0^{a/\Delta}(a-x\Delta)^2dx$, which are the stated bounds.
\end{proof}

Without holdover checks, the model is invariant under shifts in time, and a longer window never makes the attacker worse off.

\begin{lemma}[Monotonicity without holdover]\label{lem:mono}
Let~\ref{as:lin}--\ref{as:ss} hold and $J_o=0$. Then $\Es_\lambda(N+1)\le\Es_\lambda(N)$ for every $N\in\Nat$.
\end{lemma}

\begin{proof}
If $\A_N=\emptyset$ there is nothing to prove. Otherwise, let $\tau\in\A_N$ and define the delayed profile $\tau'$ for a window of $N+1$ frames by $\tau'_{i,0}=\tau'_{i,1}=0$ and $\tau'_{i,k+1}=\tau_{i,k}$ for $k\in\{1,\dots,N\}$. It satisfies (R) and (S), so $\tau'\in\A_{N+1}$, and its signature vanishes in frame $0$. By (F1), the evidence of $\tau'$ is that of the record up to frame $0$, which is zero, plus that of frames $1,\dots,N$ given the record up to frame $0$. Since $J_o=0$, the model is time invariant, and by Lemma~\ref{lem:kalman}(b) and Assumption~\ref{as:ss}, the prediction covariance of $x_{i,1}$ given the record up to frame $0$ is again $P_i=\Ric_i(P_i)$. By Theorem~\ref{thm:evidence}(a), the second term is thus the evidence of $\tau$ in a window of $N$ frames. Hence $\Es_\lambda(N+1)\le\Ev_\lambda(\tau')=\Ev_\lambda(\tau)$ for every $\tau\in\A_N$.
\end{proof}

With holdover checks, time invariance fails, because the oscillator offset accumulates from the audit on, as $\delta_ikT_s$ in~\eqref{eq:holdover}. This is why Definition~\ref{def:safe} requires safety for every shorter period.

\section{The Defender's Design}\label{sec:design}

The defender anticipates the stealthiest attack and maximizes $\Es_\lambda(N)$ over its affordable designs. We assume $J\ge1$, since otherwise there is nothing to choose.

\begin{theorem}[Design]\label{thm:design}
Let~\ref{as:lin}--\ref{as:ss} hold and $N\in\Nat$.
\begin{enumerate}
\item For every $\tau\in\R^{2N}$, $\lambda\mapsto\Ev_\lambda(\tau)$ is concave, nondecreasing in each component and continuous on $(0,\infty)^J$.
\item If $2N\Delta\ge\tc$, then $\lambda\mapsto\Es_\lambda(N)$ is concave, nondecreasing and continuous on $(0,\infty)^J$. Its maximum over $\Dset$ is attained, and maximizing it is a convex program.
\item If $2N\Delta\ge\tc$, then $\max_{\lambda\in\Dset}\min_{\tau\in\A_N}\Ev_\lambda(\tau)=\min_{\tau\in\A_N}\max_{\lambda\in\Dset}\Ev_\lambda(\tau)$, and $\Gamma_N$ has a saddle point.
\end{enumerate}
\end{theorem}

\inwords A more precise monitor never hurts, but its returns diminish; the best design can be found by convex optimization; and the defender loses nothing by announcing its design, because the max-min and min-max values coincide. A choice among discrete oscillator classes, such as TCXO, OCXO or Rb, is made by enumeration.

\begin{proof}
(1) The design enters the prediction covariance $P_r$ through $\sigma_e$, so we first remove Assumption~\ref{as:ss}. Fix $\Pi_r,\Pi_g\in\Sym^2_{++}$ that do not depend on $\lambda$, let $x_{i,-L}\sim\N(\bar x_i,\Pi_i)$, and let $\Ev^L_\lambda(\tau)$ be the resulting evidence. The proof of Theorem~\ref{thm:evidence}(a) applies with $P_i$ replaced by $\Ric_i^L(\Pi_i)$ (Lemma~\ref{lem:kalman}(b)), so $\Ev^L_\lambda(\tau)=\frac12\tau^\top G^\top\Psi_L^{-1}G\tau$ with $\Psi_L\to\Psi$ as $L\to\infty$, and hence $\Ev^L_\lambda(\tau)\to\Ev_\lambda(\tau)$. On the other hand, the proofs of Theorem~\ref{thm:evidence}(a)--(b) use only that the nuisance and the noise are independent Gaussian vectors with positive definite covariances. Applied to the whole record from frame $-L$, with the nuisance $\bar\eta$ formed by $x_{i,-L}-\bar x_i$, $w_{-L},\dots,w_{N-2}$, $d_{-L},\dots,d_{N-2}$ and $\delta$, and with the noises of all frames, they give
\[
2\Ev^L_\lambda(\tau)=\min_{\bar\eta}\big\{\|\bar G\tau-\bar F\bar\eta\|^2_{\bar R^{-1}}+\|\bar\eta\|^2_{\bar P^{-1}}\big\}
\]
for the corresponding matrices $\bar F,\bar G,\bar P,\bar R$. The matrices $\bar F$, $\bar G$ and $\Pi_i$ do not depend on $\lambda$, and every other entry of the block-diagonal matrices $\bar R^{-1}$ and $\bar P^{-1}$ is either a fixed positive number or a component $\lambda_j=\sigma_j^{-2}$ of $\lambda$. For fixed $\bar\eta$, the bracket is therefore affine and nondecreasing in $\lambda$, and its minimum over $\bar\eta$ is concave and nondecreasing. Pointwise limits preserve both properties, so $\Ev_\lambda(\tau)$ is concave and nondecreasing in $\lambda$. A concave function that is finite on an open convex set is continuous there~\cite[Thm.~10.1]{rockafellar1970}.
(2) By Theorem~\ref{thm:br}(1), $\Es_\lambda(N)=\min_{\tau\in\A_N}\Ev_\lambda(\tau)$ is finite, and a pointwise minimum of concave nondecreasing functions is concave and nondecreasing; continuity follows as in (1). A continuous function attains its maximum on the compact set $\Dset$, and maximizing a concave function over a convex set is a convex program.
(3) The function $(\lambda,\tau)\mapsto\Ev_\lambda(\tau)$ is concave and continuous in $\lambda$ by (1), and convex and continuous in $\tau$ by Theorem~\ref{thm:evidence}(a); $\Dset$ and $\A_N$ are convex and compact. Sion's minimax theorem~\cite{sion1958} gives the equality. A maximizer $\lambda^\star$ of $\Es_\lambda(N)$ exists by (2), and a minimizer $\tau^\star$ of the lower semicontinuous function $\tau\mapsto\max_{\lambda\in\Dset}\Ev_\lambda(\tau)$ exists on the compact set $\A_N$. With $v$ the common value, $\Ev_\lambda(\tau^\star)\le v=\Es_{\lambda^\star}(N)\le\Ev_{\lambda^\star}(\tau)$ for all $\lambda\in\Dset$ and $\tau\in\A_N$; taking $\lambda=\lambda^\star$ and $\tau=\tau^\star$ shows $\Ev_{\lambda^\star}(\tau^\star)=v$, so $(\lambda^\star,\tau^\star)$ is a saddle point.
\end{proof}

The design problem of Definition~\ref{def:safe} asks for the longest safe period, not for the largest evidence at a fixed period. It reduces to a sequence of convex problems.

\begin{corollary}[Designing for the audit period]\label{cor:feas}
Let~\ref{as:lin}--\ref{as:ss} hold and $h\in(0,\infty)$. For $N\in\Nat$, the set $\Dset_h(N)=\{\lambda\in\Dset:\ \text{audits every $N$ frames are $h$-safe for }\lambda\}$ is convex and compact, and $\Dset_h(N+1)\subseteq\Dset_h(N)$. Hence $\sup_{\lambda\in\Dset}T^\star_h(\lambda)=T_s\cdot\sup\{N\in\Nat:\Dset_h(N)\neq\emptyset\}$. Given $\bar N\in\Nat$ with $\Dset_h(\bar N)=\emptyset$, this supremum can be found by bisection over $N\in\{1,\dots,\bar N\}$, each step of which is a convex feasibility problem.
\end{corollary}

\begin{proof}
$\Dset_h(N)$ is the intersection over $N'\le N$ of the sets $\{\lambda\in\Dset:\Es_\lambda(N')\ge h\}$. Each is $\Dset$ itself if $2N'\Delta<\tc$, and otherwise the intersection of $\Dset$ with a superlevel set of a concave continuous function (Theorem~\ref{thm:design}(2)); both are convex and compact. The sets are nested by definition. By Definition~\ref{def:safe}, $T^\star_h(\lambda)\ge NT_s$ if and only if $\lambda\in\Dset_h(N)$, which gives the formula; nestedness makes bisection valid.
\end{proof}

Each step of the bisection must test membership in $\Dset_h(N)$ itself, that is, $\Es_\lambda(N')\ge h$ for every $N'\le N$: monotonicity of $\Es_\lambda$ in $N$ with holdover checks has not been established here (Lemma~\ref{lem:mono} covers only $J_o=0$), so testing $\Es_\lambda(N)\ge h$ alone is not justified. The upper bracket $\bar N$ can come from an explicit attack, as in Theorem~\ref{thm:cert}(c) below. If, for some $\lambda\in\Dset$, a lower bound on $\Es_\lambda(N)$ that does not depend on $N$ exceeds $h$, then $\lambda\in\Dset_h(N)$ for every $N$, the supremum is infinite, and no bisection is needed.

Where should a better oscillator go? Theorem~\ref{thm:br}(3) gives a clean answer when the per-clock coefficients $\kappa_i$ depend on the design in the simplest way.

\begin{corollary}[Where the better oscillator goes]\label{cor:alloc}
Let $\kappa^0_r,\kappa^0_g,c,\bar\lambda\in(0,\infty)$, and for $\lambda_r,\lambda_g\in[0,\infty)$ let $\kappa_i=\kappa^0_i+c\lambda_i$ and $\phi(\lambda_r,\lambda_g)=(\tc)^2/(1/\kappa_r+1/\kappa_g)$. Over $\{\lambda_r+\lambda_g=\bar\lambda,\ \lambda_r,\lambda_g\ge0\}$, $\phi$ is maximized by $\kappa_r=\kappa_g$ if $|\kappa^0_r-\kappa^0_g|\le c\bar\lambda$, and otherwise by giving all of $\bar\lambda$ to the clock with the smaller $\kappa^0_i$. When $\kappa^0_r\neq\kappa^0_g$, the clock with the smaller $\kappa^0_i$ receives more precision; when they are equal, the split is equal.
\end{corollary}

\begin{proof}
The sum $K=\kappa_r+\kappa_g=\kappa^0_r+\kappa^0_g+c\bar\lambda$ is fixed, and $\phi=(\tc)^2\kappa_r(K-\kappa_r)/K$ is a concave function of $\kappa_r$ with its maximum at $\kappa_r=K/2$. The feasible values of $\kappa_r$ form the interval $[\kappa^0_r,\kappa^0_r+c\bar\lambda]$, which contains $K/2$ if and only if $|\kappa^0_r-\kappa^0_g|\le c\bar\lambda$; otherwise the endpoint nearest to $K/2$ is optimal. At $\kappa_r=\kappa_g$, $\lambda_r-\lambda_g=(\kappa^0_g-\kappa^0_r)/c$, which is positive if $\kappa^0_r<\kappa^0_g$.
\end{proof}

\inwords $\kappa^0_i$ is what the PMU of side $i$ reveals per unit squared bias, and $c\lambda_i$ is what its holdover check adds. The hypothesis $\kappa_i=\kappa^0_i+c\lambda_i$ holds approximately for ramps, with $c=1/(2T^2)$, by Proposition~\ref{prop:regimes}(c) when $\sigma_o^2/S_N\ll\sigma_i^2$; the numbers of Section~\ref{sec:results} do not rely on it. By Proposition~\ref{prop:regimes}(a),(b), $\kappa^0_r\le f_0^2/(2s_r^2T^2)$ while $\kappa^0_g\approx f_0^2/(2q_dT_sT)$, so $\kappa^0_r\ll\kappa^0_g$ for long windows. Under this approximation, the island PMU already watches the island clock, and the better oscillator belongs on the utility side; Section~\ref{sec:results} confirms this numerically for the case study, without the approximation. Geometrically, the optimum is the rectangle of largest area for a given perimeter. The conclusion rests on the model, in particular on the white load of Assumption~\ref{as:gauss} (Section~\ref{sec:conclusion}).

\section{Case Study: A 50\,Hz Island}\label{sec:results}

\begin{figure*}[!t]
\centering
\includegraphics[width=\textwidth]{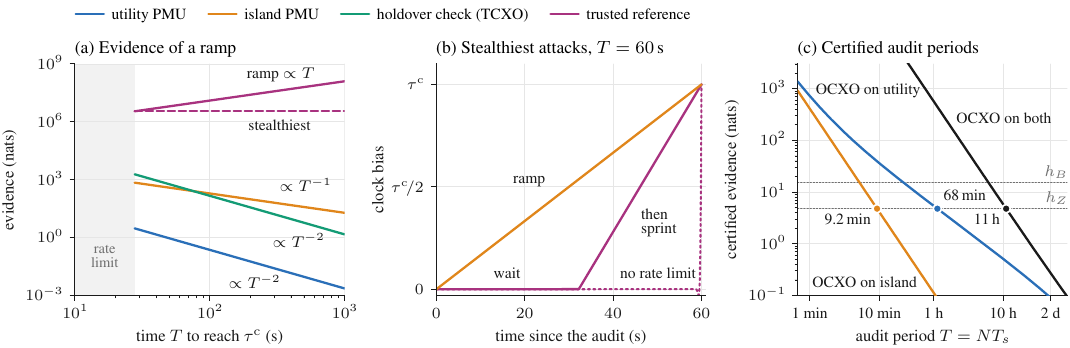}
\caption{Case study at 50\,Hz; the legend gives the color of each monitor, as in Fig.~\ref{fig:system}. (a)~Evidence of a ramp that reaches $\tc$ on one clock in $T$ seconds, for each monitor alone, with the slopes of Proposition~\ref{prop:regimes} and Table~\ref{tab:regimes}. One clock cannot reach $\tc$ in less than $\tc/\gb=27.8$\,s (shaded). Against the trusted reference, the stealthiest attack (dashed, computed by the QP) reveals numerically the same evidence for every $T\ge27.8$\,s. (b)~Stealthiest profiles in a 60\,s window, computed by the QP of Theorem~\ref{thm:br}: numerically close to a ramp against the island PMU with a TCXO holdover check, and to wait-then-sprint against the trusted reference; without the rate limit, the attacker would jump in the last frames (dotted). (c)~Certified least evidence $E_{\rm low}(N)$ of a successful attack (Theorem~\ref{thm:cert}) against the audit period $T=NT_s$, for three oscillator placements. Each curve crosses the threshold $h_Z$ at its certified safe period (dots). The curve for TCXOs on both clocks (not shown) nearly coincides with ``OCXO on island'' (8.7 versus 9.2\,min).}
\label{fig:results}
\end{figure*}

\emph{Setup.} We use the values in the last column of Table~\ref{tab:symbols}: $f_0=50$\,Hz, $T_s=20$\,ms, $T_g=0.5$\,s, $\sigma_\theta=\sigma_e=10^{-3}$\,rad, $\sigma_\nu=\sqrt{q_d}=1$\,mHz, $\sqrt{q_w}=0.1$\,mHz, $\sigma_o=20$\,ns, $\tc=0.556$\,ms ($10^\circ$) and $\gb=2\times10^{-5}$. The selected PMU noise standard deviations are substantially smaller than the steady-state limits of~\cite{c37118}, a total vector error of $1\%$ (about $0.01$\,rad) and a frequency error of 5\,mHz, and $T_s$ is its 50\,frames/s reporting period. The rate limit is deliberately pessimistic: it is six times the fastest drag demonstrated against a PMU and fifteen times the loss-of-lock limit of a power-grid receiver (Section~\ref{sec:clock}). Since $\A_N$ grows with $\gb$, a larger $\gb$ can only shorten the safe periods. With $\gb=3.3\times10^{-6}$ instead, every period of Table~\ref{tab:audit} that involves an oscillator or the reference changes by less than $0.5\%$, and the period without monitors grows from 26\,s ($h_Z$) and 24\,s ($h_B$) to $1.4$\,min, the time that two clocks then need to reach $\tc$. A TCXO, an OCXO and a rubidium standard are rated at $10^{-6}$, $10^{-8}$ and $10^{-10}$, within the frequency offsets of~\cite[Table~17.5]{lombardi2002}. The source specifies no distribution. As a calibration assumption of ours, we read each rating as $z_{1-\alpha}$ standard deviations of a zero-mean Gaussian offset, with $z_{1-\alpha}=\Phi^{-1}(1-\alpha)=3.090232$ for $\alpha=10^{-3}$: a fraction $\alpha$ of the oscillators exceed the rating in each direction, $2\alpha$ in absolute value. Hence $\sigma_i$ is the rating divided by $z_{1-\alpha}$, displayed rounded as $3.236\times10^{-7}$, $3.236\times10^{-9}$ and $3.236\times10^{-11}$ and used unrounded in every computation, as are $h_Z=\frac12z_{1-\alpha}^2$ and $h_B$. Aging, which the model neglects, is at most $5\times10^{-7}$, $5\times10^{-9}$ and $10^{-10}$ per year~\cite{lombardi2002}. It is negligible over the periods of Table~\ref{tab:audit}, except for two rubidium standards, whose 45-day period lets aging reach about $40\%$ of $\sigma_i$; that entry, marked in Table~\ref{tab:audit}, is indicative only. Solving the Riccati equations gives $s_r=0.24$\,mHz, $s_g=0.10$\,mHz, $\sqrt{(P_g)_{22}}=0.12$\,mHz and $N_g=136$ frames. The utility PMU therefore reveals no more than an oscillator of spread $s_r/f_0=4.8\times10^{-6}$, fifteen times worse than a TCXO. The PMU parts of the record are evaluated exactly through their steady-state Kalman innovations, and the stealthiest attacks by an active-set QP solver. The script \texttt{code/evidence\_island.py} of the supplementary material regenerates every number, Fig.~\ref{fig:results} and Table~\ref{tab:brackets}.

\emph{The three regimes.} A ramp reaching $\tc$ in 60\,s reveals $0.64$\,nats to the utility PMU (Proposition~\ref{prop:regimes}(a): at most $1.83$), $409$ to a TCXO holdover check, $323$ to the island PMU (Lemma~\ref{lem:bounds}(b): between $311$ and $332$) and $7.6\times10^6$ to the trusted reference (Proposition~\ref{prop:regimes}(d): at least $7.6\times10^6$). Fig.~\ref{fig:results}(a) shows the slopes of Table~\ref{tab:regimes} over two decades of $T$. Note the crossover near 80\,s: for slower attacks, the island PMU reveals more than a TCXO.

\emph{Stealthiest attacks.} Against the island PMU with a TCXO, the QP solution is within $4\times10^{-4}\tc$ of the ramp and reveals $732$\,nats. Against the trusted reference, it sprints at $\gb$ in the last $27.8$\,s and reveals $3.52\times10^6$\,nats for windows of 30, 40 and 60\,s alike, which numerically agrees with the lower bound of Proposition~\ref{prop:shape}(b) at the reported precision; the 60\,s ramp reveals $7.61\times10^6$. Without the rate limit the attacker would reveal only $4.98\times10^4$, so the rate limit multiplies its cost by 71. With a TCXO on both clocks, the two-clock QP puts $64.1\%$ of the bias on the utility clock and reveals $262.8$\,nats; the harmonic rule~\eqref{eq:harmonic}, with the $\kappa_i$ of ramps, predicts $64.1\%$ and $262.9$\,nats.

\emph{Certified audit periods.} Computing $T^\star_h(\lambda)$ exactly would require the QP of Theorem~\ref{thm:br} for every $N$ up to millions of frames. Instead, we bracket it.

\begin{theorem}[Certified audit periods]\label{thm:cert}
Let~\ref{as:lin}--\ref{as:ss} hold, $\tc>2\Delta$, and fix a design and $h\in(0,\infty)$. Let $\mathbf 1_{o_r}=\mathbf 1_{r\in\Jset}$, $\mathbf 1_{o_g}=\mathbf 1_{g\in\Jset}$ and $\mathbf 1_e=\mathbf 1_{e\in\Jset}$ indicate the installed monitors, and for $N\in\Nat$ with $N\ge2$ and $a\in[0,\infty)$ let
\begin{align*}
\chi_g(N)&=\frac{\omega_0^2}{2V^g_N}+\frac{\mathbf 1_{o_g}}{2\big(\sigma_g^2(N-1)^2T_s^2+\sigma_o^2\big)},\\
\chi_r(N)&=\frac{\mathbf 1_{o_r}}{2\big(\sigma_r^2(N-1)^2T_s^2+\sigma_o^2\big)},\\
\mathcal L_g(a;N)&=\chi_g(N)(a-\Delta)_+^2,\\
\mathcal L_r(a;N)&=\chi_r(N)(a-\Delta)_+^2+\frac{\mathbf 1_e\,\omega_0^2}{2\sigma_{\theta e}^2}\sum_{k\ge1}(a-k\Delta)_+^2,
\end{align*}
with $\mathcal L_i(a;N)=+\infty$ if $a>N\Delta$. Let $E_{\rm low}(N)=\min_{a\in[0,\tc]}\{\mathcal L_r(a;N)+\mathcal L_g(\tc-a;N)\}$. Then:
\begin{enumerate}
\item[(a)] $\Es(N)\ge E_{\rm low}(N)$, and $E_{\rm low}$ is nonincreasing in $N$;
\item[(b)] $T^\star_h\ge T_{\rm low}:=T_s\cdot\sup\{N\ge2:E_{\rm low}(N)\ge h\}$;
\item[(c)] if some $\tau\in\A_{N_1}$ has $\Ev(\tau)<h$, then $T^\star_h<N_1T_s$.
\end{enumerate}
\end{theorem}

\begin{proof}
(a) Let $\tau\in\A_N$, $a_r=\tau_{r,N}$ and $a_g=-\tau_{g,N}$, so $a_r+a_g=\tc$ and $|a_i|\le N\Delta$. By (R), $|\tau_{i,N-1}|\ge|a_i|-\Delta$. By Theorem~\ref{thm:evidence}(d) and Lemma~\ref{lem:bounds}(b),(c), $\Ev_{y_g}(\tau_g)+\Ev_{o_g}(\tau_g)\ge\chi_g(N)(|a_g|-\Delta)_+^2$ and $\Ev_{o_r}(\tau_r)\ge\chi_r(N)(|a_r|-\Delta)_+^2$. If the reference is installed, the proof of Proposition~\ref{prop:shape}(b) gives $\Ev_{y_r}(\tau_r)\ge\omega_0^2\sum_{k\ge1}(|a_r|-k\Delta)_+^2/(2\sigma_{\theta e}^2)$; otherwise we use $\Ev_{y_r}\ge0$, which only weakens the bound. Hence $\Ev(\tau)\ge\mathcal L_r(|a_r|;N)+\mathcal L_g(|a_g|;N)$. Each $\mathcal L_i(\cdot;N)$ is nondecreasing on $[0,\infty)$. If $a_r<0$, then $|a_g|>\tc$, and replacing $(|a_r|,|a_g|)$ by $(0,\tc)$ does not increase the sum; similarly if $a_g<0$. So $\Ev(\tau)\ge E_{\rm low}(N)$ for every $\tau\in\A_N$. For monotonicity, each term of $V^g_N$ is nondecreasing in $N$, so $\chi_g$ and $\chi_r$ are nonincreasing; the sum over $k$ does not depend on $N$; and the set where $\mathcal L_i=+\infty$ shrinks as $N$ grows.
(b) Since $\tc>2\Delta$, $\A_1=\emptyset$ and $\Es(1)=+\infty$. By (a), the set $\{N\ge2:E_{\rm low}(N)\ge h\}$ is of the form $\{2,\dots,N_{\rm low}\}$ (or all $N\ge2$), and on it $\Es(N)\ge E_{\rm low}(N)\ge h$. So audits every $N_{\rm low}=T_{\rm low}/T_s$ frames are $h$-safe.
(c) $\Es(N_1)\le\Ev(\tau)<h$, so no period of $N\ge N_1$ frames is $h$-safe.
\end{proof}

\emph{Computing $E_{\rm low}$.} A minimum over a grid of splits can overstate $E_{\rm low}(N)$, so we compute it exactly. In the utility share $a$, the objective is convex and continuously differentiable on the feasible interval $[\max\{0,\tc-N\Delta\},\min\{\tc,N\Delta\}]$. Without the reference, and when no boundary binds, its minimizer is $a=\Delta+\chi_g(N)(\tc-2\Delta)/\big(\chi_r(N)+\chi_g(N)\big)$, and $E_{\rm low}(N)=(\tc-2\Delta)^2\chi_r(N)\chi_g(N)/\big(\chi_r(N)+\chi_g(N)\big)$. In general, we bracket the zero of the derivative by bisection and take the tangent at the left end of the final bracket, which by convexity never exceeds the minimum; a relative margin of $10^{-12}$ absorbs rounding. Since $E_{\rm low}$ is nonincreasing, $T_{\rm low}$ then follows by bisection over whole frames.

\begin{table}[t]
\caption{Safe audit period: the per-check rule versus the certified lower bound $T_{\rm low}$ on $T^\star_h$ (Theorem~\ref{thm:cert}(b)), which fuses all channels; $T^\star_h$ exceeds it by less than $0.6\%$ (Table~\ref{tab:brackets}). At $h_Z$ ($\alpha=10^{-3}$) the matched test, and at $h_B$ (one false alarm per day) the matched SR detector, detects every successful attack with probability at least $\frac12$}\label{tab:audit}
\centering\footnotesize
\setlength{\tabcolsep}{4pt}
\begin{tabular}{@{}llrrr@{}}
\toprule
Utility $r$ & Island $g$ & Per-check & Fused, $h_Z$ & Fused, $h_B$\\
\midrule
-- & -- & 14\,s & 26\,s & 24\,s\\
TCXO & TCXO & 4.1\,min & 8.7\,min & 4.7\,min\\
TCXO & OCXO & 8.1\,min & 9.2\,min & 5.2\,min\\
OCXO & TCXO & 8.1\,min & 68\,min & 22\,min\\
OCXO & OCXO & 6.9\,h & 11.0\,h & 6.1\,h\\
Rb & OCXO & 13.6\,h & 16.0\,h & 8.8\,h\\
Rb & Rb$^\dagger$ & 28.5\,d & 45.4\,d & 25.4\,d\\
\midrule
trusted ref. & TCXO & -- & 68\,min & 22\,min\\
trusted ref. & OCXO & -- & 16.0\,h & 8.8\,h\\
\bottomrule
\end{tabular}

\smallskip
\parbox{\columnwidth}{\footnotesize $^\dagger$Indicative only: over this period, oscillator aging, which the model neglects, reaches about $40\%$ of $\sigma_i$.}
\end{table}

\begin{table}[t]
\caption{Brackets $T_{\rm low}\le T^\star_h\le T_{\rm up}$ behind Table~\ref{tab:audit}, in seconds, both whole numbers of frames: $T_{\rm low}$ from Theorem~\ref{thm:cert}(b), $T_{\rm up}=(N_1-1)T_s$ from an explicit attack of $N_1$ frames (Theorem~\ref{thm:cert}(c)); gap $=(T_{\rm up}-T_{\rm low})/T_{\rm low}$. Without monitors, $T^\star_h$ is exact}\label{tab:brackets}
\centering\scriptsize
\setlength{\tabcolsep}{2.4pt}
\begin{tabular}{@{}llrrrrrr@{}}
\toprule
& & \multicolumn{3}{c}{$h_Z$} & \multicolumn{3}{c}{$h_B$}\\
\cmidrule(lr){3-5}\cmidrule(l){6-8}
$r$ & $g$ & $T_{\rm low}$ & $T_{\rm up}$ & gap & $T_{\rm low}$ & $T_{\rm up}$ & gap\\
\midrule
-- & -- & 26.08 & 26.08 & 0 & 24.28 & 24.28 & 0\\
TCXO & TCXO & 522.34 & 523.62 & 0.25\% & 281.40 & 282.18 & 0.28\%\\
TCXO & OCXO & 554.74 & 555.98 & 0.22\% & 310.12 & 310.80 & 0.22\%\\
OCXO & TCXO & 4079.70 & 4093.54 & 0.34\% & 1327.10 & 1333.02 & 0.45\%\\
OCXO & OCXO & 39720.98 & 39778.98 & 0.15\% & 22084.66 & 22116.74 & 0.15\%\\
Rb & OCXO & 57523.04 & 57612.28 & 0.16\% & 31646.40 & 31694.68 & 0.15\%\\
Rb & Rb & 3923217.56 & 3928876.04 & 0.14\% & 2193056.40 & 2196219.26 & 0.14\%\\
ref. & TCXO & 4098.96 & 4115.34 & 0.40\% & 1327.40 & 1335.12 & 0.58\%\\
ref. & OCXO & 57507.50 & 57612.28 & 0.18\% & 31622.28 & 31694.68 & 0.23\%\\
\bottomrule
\end{tabular}
\end{table}

The fused columns of Table~\ref{tab:audit} report $T_{\rm low}$ of Theorem~\ref{thm:cert}(b), a period that is safe against every attack. For every design with a monitor, explicit feasible attacks give an upper bound $T_{\rm up}$ on $T^\star_h$ by Theorem~\ref{thm:cert}(c): ramps on the two clocks with the best split of the bias, delayed to end at the audit when a clock has no oscillator. Their evidence is evaluated in closed form, for any $N$ and without extrapolation. Driven by a ramp, the steady-state innovation filter of each PMU has an affine particular solution plus a geometrically decaying transient, so the sum of the squared whitened innovations follows from power sums, geometric matrix sums and a Stein equation; up to 50\,min it agrees with direct summation to a relative $5\times10^{-12}$. Let $N_1$ be the first frame count found at which such an attack has evidence below $h$. By Theorem~\ref{thm:cert}(c), $T^\star_h<N_1T_s$, so on the frame lattice $T^\star_h\le T_{\rm up}:=(N_1-1)T_s$. The lower endpoint $T_{\rm low}$ is computed in whole frames with the exact $E_{\rm low}$ described after Theorem~\ref{thm:cert}. Table~\ref{tab:brackets} lists both endpoints; the relative gap is below $0.6\%$ throughout, a numerical statement for these designs.

Without monitors, we solve the two-clock QP of Theorem~\ref{thm:br} by bisection over $N$, which Lemma~\ref{lem:mono} makes valid, and certify the crossing by weak duality. At 1304 frames, a Lagrange dual point shows $\Es\ge4.786>h_Z$, and at 1305 frames a feasible attack has evidence $4.710<h_Z$; at $h_B$, the values are $15.291$ at 1214 frames and $15.131$ at 1215. The primal and dual values agree to nine digits, so $T^\star_{h_Z}=26.08$\,s and $T^\star_{h_B}=24.28$\,s exactly.

Fig.~\ref{fig:results}(c) shows how the periods arise. For each design, the certified evidence $E_{\rm low}(N)$ decays with the period, and the safe period is where it falls below the threshold, as in Corollary~\ref{cor:feas}. Moving the OCXO from the island to the utility side lifts the curve by more than an order of magnitude at long periods, and the crossing of $h_Z$ moves from 9.2 to 68\,min.

\emph{The per-check rule.} The column ``Per-check'' is a baseline rule that treats each holdover check as a separate alarm. Check $i$ flags drifts above $z_{1-\alpha/4}\sigma_i=3.48\sigma_i$ (a two-sided test with $\alpha=10^{-3}$ split between two checks). Let $b_i=\min\{\gb,3.48\sigma_i\}$ for a clock with an oscillator and $b_i=\gb$ for one without. An attacker that drifts just below both limits, in opposite directions, needs the time $T_{\rm pc}=\tc/(b_r+b_g)$, and the rule deems shorter periods safe. For the oscillators of Table~\ref{tab:audit}, $3.48\sigma_i<\gb$, so $b_i=3.48\sigma_i$.

The two kinds of column do not apply the same statistical criterion. The per-check rule runs two separate two-sided tests with a split false-alarm budget, each blind to the phasor data and to the attack profile. The threshold $h_Z$ refers to one one-sided test that uses all channels and is matched to a signed, fully specified profile. The gains at $h_Z$ therefore combine fusion, phasor information, sidedness and knowledge of the alternative; the column $h_B$ removes the knowledge of the onset. Separating these effects requires a fused composite detector calibrated to the same false-alarm and detection requirements, which we leave to future work.

\emph{Why fusing helps.} Within the evidence criterion, three effects explain the gains at $h_Z$:
\begin{itemize}
\item \emph{Fusing the holdover checks.} Against one combined test, the attacker can hide a drift of order $(\sigma_r^2+\sigma_g^2)^{1/2}$ rather than $\sigma_r+\sigma_g$. With equal oscillators, this lengthens the period by the factor $2\cdot3.48/(\sqrt2\,z_{1-\alpha})=3.48/\sqrt{h_Z}=1.59$ (row Rb/Rb), of which $\sqrt2$ is due to fusion and $3.48/3.09$ to the one-sided matched threshold.
\item \emph{Using the phasor data.} The island PMU's evidence of a ramp decays as $1/T$ and a TCXO's as $1/T^2$, so beyond about 80\,s the island PMU dominates, and the TCXO/TCXO period doubles.
\item \emph{Placement.} With one OCXO and one TCXO, the rule gives 8.1\,min either way. Fused evidence gives 68\,min with the OCXO on the utility side, but only 9.2\,min with it on the island side, as Corollary~\ref{cor:alloc} predicts.
\end{itemize}
A trusted utility reference pushes nearly all of the bias onto the island clock, so the period equals that of a near-perfect utility oscillator. The threshold $h_B$ shortens every period but preserves the ranking of the designs. For a budget of two TCXOs' worth of precision, $\lambda_r+\lambda_g=2\sigma_{\rm TCXO}^{-2}$, we maximize the certified period $T_{\rm low}$, which respects every shorter horizon as Definition~\ref{def:safe} requires, over the utility shares $0.01,0.02,\dots,1$ of the budget; the share $1$ leaves the island clock without an oscillator ($\lambda_g=0$). The best share on this grid is $1$, which lengthens $T_{\rm low}$ from 8.7 to 11.9\,min at $h_Z$ ($T_{\rm up}=11.9$\,min) and from 4.7 to 6.1\,min at $h_B$. This is a numerical best over specified hardware designs, not the optimum of a continuous design problem. Our theorems assume positive precisions on a fixed set of monitors, and $\lambda_g=0$ is not the limit of small positive $\lambda_g$: a holdover check of vanishing precision still sees deviations of the clock from a straight line, which a removed one does not, so $T^\star_h$ may jump at $\lambda_g=0$. The brackets bound any such jump. For the share $0.999$, $T^\star_{h_Z}$ lies in $[711.38,712.96]$\,s, and for the share $1$ in $[711.68,713.26]$\,s.

\emph{Robustness.} Two checks probe the conclusions beyond the nominal parameters and beyond Assumption~\ref{as:lin}.

\emph{Load variability.} The placement gain is driven by the island PMU's evidence, which scales as $1/q_d$ (Proposition~\ref{prop:regimes}(b)). With the load standard deviation scaled by $0.3$, $1$ and $3$, the certified periods $T_{\rm low}$ with the OCXO on the utility versus the island side are 514.6 versus 9.2\,min, 68.0 versus 9.2\,min and 13.7 versus 9.2\,min at $h_Z$, gains of $55.7$, $7.4$ and $1.5$; at $h_B$, the gains are $38.4$, $4.3$ and $1.3$. The direction of the placement rule holds throughout; its size is set by the load variability, so $q_d$ should be estimated before deployment.

\emph{State-dependent terms.} Equation~\eqref{eq:pmu} omits the terms $-2\pi\nu_{i,k}\tau_{i,k}$ in the reported phase and $-\nu_{i,k}\gamma_{i,k}$ in the reported frequency. With them, the signature depends on the state, the attacked record's covariance differs from the unattacked one, and, because the controller shapes $\nu_{g,k}$, the evidence depends on the controller: the dual effect that Theorem~\ref{thm:evidence}(a) excludes. For a fixed profile the closed loop under the PI law~\eqref{eq:ctrl} remains linear in the state and the noises, so the record is still Gaussian and the exact divergence follows from its mean and covariance. With the priors of Assumption~\ref{as:ss}, for opposite ramps on the two clocks reaching $\tc$ in 14\,s (drift $0.99\gb$), the exact evidence differs from~\eqref{eq:quad} by $-7.7\times10^{-5}$ (tuning A) and $-7.9\times10^{-5}$ (tuning B) in relative terms, of $356.66$\,nats; the covariance change contributes at most $7\times10^{-5}$\,nats. For a utility-clock sprint at $\gb$ to $\tc/2$ against the trusted reference, the relative difference is $2\times10^{-9}$ under both tunings. The dual effect exists but is four orders of magnitude below the evidence, because the omitted terms are smaller than the retained ones by the factor $|\nu_{i,k}|/f_0$.

\section{Conclusion}\label{sec:conclusion}

We posed the defense of a synchronous island against GNSS timing attacks as an inspection game, in which the defender chooses its monitors, the attacker drags two clocks to an unsafe bias between audits, and the payoff is the evidence. One identity carries the analysis: the evidence is the cost of the cheapest innocent explanation of what the defender sees. It tells which monitors work (those that cannot blame a drift on normal operation), why ramps are nearly optimal in value against the island PMU, whose innocent explanation must be paid for in every frame, and why late sprints come within a bounded factor of the optimum against a reference that penalizes squared bias in every frame (the case-study optima have these shapes numerically), and, under the stated model, where a better oscillator belongs (where the PMU sees least). The attacker's problem is a strictly convex QP, the defender's is a convex program, and the game has a saddle point. Because explanations bound the evidence from above and statistics bound it from below, safe audit periods can be certified rather than merely estimated, in the sense that every successful attack leaves at least the threshold of evidence.

Three lessons follow for practice, within the model and the case study. \emph{Fuse:} combining existing channels increases the worst-case KL evidence; in the case study, the resulting evidence-based audit bounds are up to 8.4 times the per-check baseline at $h_Z$, with no new hardware. \emph{Place:} the better oscillator belongs on the utility side, because the island PMU already watches the island clock. \emph{Anchor:} a trusted phase reference prevents waiting from hiding a fixed nonzero utility-clock bias, makes the utility clock expensive to drag, and pushes the attack toward the island clock.

The results rest on the model, and the load model matters most: if the load were an unknown constant rather than white noise, the island PMU would explain a drift once, and the placement gain would shrink, so $q_d$ should be estimated before deployment. Attacks are open loop, audits are ideal, success is defined by the differential bias, a proxy for the phase at reconnection that is accurate to within the tracking margin of Section~\ref{sec:damage}, the reports omit terms of relative order $2\times10^{-3}$, and oscillator aging is neglected. Most importantly for practice, the safe periods certify evidence: they guarantee detection for a test matched to each attack, not for one detector that faces an unknown profile. A fused composite detector with a calibrated guarantee for every admissible profile, adaptive attacks, controller saturation and placement across many buses are left to future work.

\end{document}